\documentclass[preprint]{aastex}

\newcommand{\be}{\begin{equation}}
\newcommand{\ee}{\end{equation}}
\newcommand{\undex}{ \beta } 
\newcommand{\angmom}{ \eta } 
\newcommand{\tratio}{ \lambda } 
\newcommand{\energy}{ {\cal E}} 
\newcommand{\tlyap}{\tau_{\rm{ly}}}
\newcommand{\tmass}{\tau} 
\newcommand{\growth}{ \Lambda } 
\def\lta{\,\raise 0.3 ex\hbox{$ < $}\kern -0.75 em
 \lower 0.7 ex\hbox{$\sim$}\,}
\def\gta{\,\raise 0.3 ex\hbox{$ > $}\kern -0.75 em
 \lower 0.7 ex\hbox{$\sim$}\,} 

\begin{document}

\title{\bf EVOLUTION OF PLANETARY SYSTEMS \\ 
WITH TIME DEPENDENT STELLAR MASS LOSS} 

\author{Fred C. Adams$^{1,2}$, Kassandra R. Anderson$^{1}$, 
and Anthony M. Bloch$^{3}$} 

\affil{$^1$Physics Department, University of Michigan, Ann Arbor, MI 48109} 

\affil{$^2$Astronomy Department, University of Michigan, Ann Arbor, MI 48109} 

\affil{$^3$Math Department, University of Michigan, Ann Arbor, MI 48109 }

\begin{abstract} 
Observations indicate that intermediate mass stars, binary stars, and
stellar remnants often host planets; a complete explanation of these
systems requires an understanding of how planetary orbits evolve as
their central stars lose mass. Motivated by these dynamical systems,
this paper generalizes in two directions previous studies of orbital
evolution in planetary systems with stellar mass loss: [1] Many
previous treatments focus on constant mass loss rates and much of this
work is carried out numerically. Here we study a class of single
planet systems where the stellar mass loss rate is time dependent. The
mass loss rate can be increasing or decreasing, but the stellar mass
always decreases monotonically. For this class of models, we develop
analytic approximations to specify the final orbital elements for
planets that remain bound after the epoch of mass loss, and find the
conditions required for the planets to become unbound.  We also show
that for some mass loss functions, planets become unbound only in the
asymptotic limit where the stellar mass vanishes. [2] We consider the
chaotic evolution for two planet systems with stellar mass loss. Here
we focus on a model consisting of analogs of Jupiter, Saturn, and the
Sun. By monitoring the divergence of initially similar trajectories
through time, we calculate the Lyapunov exponents of the system. This
analog solar system is chaotic in the absence of mass loss with
Lyapunov time $\tlyap\approx5-10$ Myr; we find that the Lyapunov time
decreases with increasing stellar mass loss rate, with a nearly linear
relationship between the two time scales. Taken together, the results
of this paper help provide an explanation for a wide range of
dynamical evolution that occurs in solar systems with stellar mass
loss.
\end{abstract}


\section{Introduction} 
\label{sec:intro} 

Solar systems orbiting other stars display a diverse set of
architectures and motivate further studies concerning the dynamics of
planetary systems. Part of the richness of this dynamical problem
arises from the intrinsic complexity of N-body systems, even in the
absence of additional forces (Murray \& Dermott 1999). The ledger of
physical behavior experienced by such systems is enormous, and
includes mean motion resonances, secular interactions, and sensitive
dependence on the initial conditions (chaos). Additional complications
arise from additional forces that are often present: During early
stages of evolution, circumstellar disks provide torques that
influence orbital elements, and turbulent fluctuations act on young
planets. Over longer time scales, solar systems are affected by tidal
forces from both stars and planets, and by general relativistic
corrections that lead to orbital precession. Another classic problem
in solar system dynamics concerns planetary orbits around central
stars that are losing mass (Gyld{\'e}n 1884, Jeans 1924; see also
Hadjidemetriou 1963, 1966).  Although this issue has received some
recent attention (see below), this paper expands upon existing work in
two main directions.  Recent work often focuses on the particular case
of constant mass loss rates, although stellar mass loss rates
typically vary with time; in addition, this recent work is primarily
carried out numerically (note that Veras et al. 2013 use numerical
simulations to consider more realistic, time-dependent stellar mass
loss). In this paper, for single planet systems, we extend existing
calculations to account for time dependence of the mass loss rates and
obtain a number of analytic results.  For systems with two or more
planets, we also show that the Lyapunov exponents, which determine the
time scales for chaotic behavior, depend on the time scales for mass
loss.  As outlined below, these two results can account for a great
deal of the possible behavior in solar systems where the central star
loses mass.

A number of previous studies have considered planetary dynamics for
host stars that are losing mass. For our own Solar System, long term
integrations have been carried out to study the fate of the planets in
light of mass loss from the dying Sun (Duncan \& Lissauer 1998).
Recent related work estimates an effective outer boundary $r_B$ to the
Solar System (due to stellar mass loss) in the range $r_B = 10^3-10^4$
AU, where orbiting bodies inside this scale remain safely bound (Veras
\& Wyatt 2012). Planets orbiting more massive stars, which lose a
larger percentage of their mass, have their survival threatened by
possible engulfment during the planetary nebula phase (Villaver \&
Livio 2007, Mustill \& Villaver 2012), and are more likely to become
unbound due to stellar mass loss alone (Villaver \& Livio 2009). In
the future of our own system, Earth is likely to be engulfed by the
Sun (Schr{\"o}der \& Connon Smith 2008), but planets in wider orbits
are expected to survive.  However, gaseous planets that escape
engulfment are still subject to evaporation and can experience
significant mass loss (Bear \& Soker 2011, Spiegel \& Madhusudhan
2012). For planets orbiting stars that are losing mass, a more general
treatment of the dynamics has been carried out for both single planet
systems (Veras et al. 2011) and multiple planet systems (Veras \& Tout
2012; Voyatzis et al. 2013); these studies provide a comprehensive
analysis for the particular case of constant mass loss rates. In 
addition to causing planets to become unbound, stellar mass loss can
drive orbital evolution that leads to unstable planetary systems
surrounding the remnant white dwarfs remaining at the end of stellar
evolution (Debes \& Sigurdsson 2002). Indeed, observations indicate
that white dwarfs can anchor both circumstellar disks (Melis et
al. 2010) and planetary systems (Zuckerman et al. 2010); many white
dwarf atmospheres contain an excess of heavy elements (Melis et
al. 2010; Jura 2003), which is assumed to be a signature of accretion
of a secondary body (a planet or asteroid).  Finally, mass loss in 
binary star systems can lead to orbital instability, allowing planets
to change their host star (Kratter \& Perets 2012; for additional
related work, see also Perets \& Kratter 2012, Moeckel \& Veras 2012).

This paper builds upon the results outlined above.  Most previous
studies have focused on stellar mass loss rates that are constant in
time, and most recent work has been carried out numerically.  However,
stars generally have multiple epochs of mass loss, the corresponding
rates are not constant, and these solar systems span an enormous range
of parameter space. It is thus useful to obtain general analytic
results that apply to a wide class of mass loss functions.  The first
goal of this paper is to study single planet systems where the stellar
mass loss rate varies with time.  As the system loses mass, the
semimajor axis of the orbit grows, and the planet becomes unbound for
critical values of the stellar mass fraction $m_f=M_f/M_{0\ast}$. For
systems that become unbound, we find the critical mass fraction $m_f$
as a function of the mass loss rate and the form of the mass loss
function. In other systems with mass loss, the orbit grows but does
not become unbound. In these cases, we find the orbital elements at
the end of the mass loss epoch, again as a function of the mass loss
rate and the form of the mass loss function.  For initially circular
orbits and slow mass loss (time scales much longer than the initial
orbital period), the critical mass fraction and/or the final orbital
elements are simple functions of parameters that describe the mass
loss rate.  For initial orbits with nonzero eccentricity, however, the
outcomes depend on the initial orbital phase. In this latter case, the
allowed values of the critical mass fraction $m_f$ (or the final
orbital elements) take a range of values, which we estimate herein.

Next we consider the effects of additional planets on the results
described above. If the planets are widely spaced, they evolve much
like individual single planet systems.  However, if the planets are
sufficiently close together so that planet-planet interactions are
important, the systems are generally chaotic. As a secondary goal,
this paper estimates the Lyapunov exponents for two-planet systems
with stellar mass loss.  For the sake of definiteness, we focus on
planetary systems containing analogs of the Sun, Jupiter, and Saturn,
i.e., bodies with the same masses and (usually) the same orbital
elements.  We find that the time scale for chaos (the inverse of the
Lyapunov exponent) is proportional to the mass loss time scale. As a
result, by the time the star has lost enough mass for the planets to
become unbound, the planets have begun to erase their initial
conditions through chaos.  For systems that evolve far enough in time,
one can use the semi-analytic results derived for single planet
systems with initially circular orbits (see above) as a rough estimate
of the conditions (e.g., the final value $\xi_f$ of the radius)
required for a planet to become unbound. Multiple planets and nonzero
initial eccentricity act to create a distribution of possible values
(e.g., for $\xi_f$) centered on these results.  Since the two-planet
systems are chaotic, and display sensitive dependence on initial
conditions, one cannot unambiguously predict the value of $\xi_f$
required for an planet to become unbound.

For completeness, we note that the problem of planetary orbits with
stellar mass loss is analogous to the problem of planetary orbits with
time variations in the gravitational constant $G$. For single planet
systems (the pure two-body problem), the gravitational force depends
only on the (single) product $GM_\ast$, so that the two problems are
equivalent (e.g., Vinti 1974). However, for the case with time varying
gravitational constant, the product $GM_\ast$ could increase with
time. Current experimental limits show that possible variations occur
on time scales much longer than the current age of the universe (see
the review of Uzan 2003), so that these effects only (possibly) become
important in the future universe (Adams \& Laughlin 1997).  We also
note that when considering time variations of the constants, one
should use only dimensionless quantities, in this case the
gravitational fine structure constant $\alpha_G=Gm_P^2/c\hbar$ 
(e.g., Duff et al. 2002).

This paper is organized as follows. We first present a general
formulation of the problem of planetary orbits with stellar mass loss
in Section \ref{sec:genmodel} and then specialize to a class of models
where the mass loss has a specific form (that given by equations
[\ref{udotcon}] and [\ref{mform}]).  These models include a wide range
of behavior for the time dependence of the mass loss rates, including
constant mass loss rates, exponential mass loss, and mass loss rates
that decrease quickly with time; these results are described in
Section \ref{sec:results}. In the following Section \ref{sec:lyapunov}, 
we consider two planet systems and calculate the Lyapunov time scales
for a representative sample of mass loss functions. Next we apply
these results to representative astronomical systems in Section
\ref{sec:apply}.  The paper concludes, in Section \ref{sec:conclude},
with a summary of our results and a discussion of their implications.

\section{Model Equations for Orbits with Stellar Mass Loss} 
\label{sec:genmodel} 

In this section we develop model equations for solar systems where the
central star loses mass.  We assume that mass loss takes place
isotropically, so that the rotational symmetry of the system is
preserved and the total angular momentum is a constant of motion.
This constraint is explicitly satisfied in the analytical solutions
that follow. For the numerical solutions, this property is used as a
consistency check on the numerical scheme.  On the other hand, the
total energy of the system is not conserved because the total mass
decreases with time (equivalently, the system no longer exhibits time
reversal symmetry).

General forms for the equations of motion with variable stellar mass
are presented in many previous papers (from Jeans 1924 to Veras et al.
2011). Some of the subtleties of the various approaches are outlined
in Hadjidemetriou (1963). In this section and the next we specialize
to systems with a single planet and focus on the case where the planet
mass is much smaller than the stellar mass, $M_p \ll M_\ast$.  The
specific angular momentum $J$ can be written in the form 
\be
J^2 = G M_{0\ast} a_0 \, \angmom \, , 
\label{amdef} 
\ee
where $a_0$ is the starting semimajor axis and $M_{0\ast}$ is the
initial mass of the star. Equation (\ref{amdef}) can be taken as the
definition of the angular momentum parameter $\angmom$. For a starting
circular orbit $\angmom$ = 1, whereas eccentric orbits have $\angmom$
= $1 - e^2 < 1$, where $e$ is the initial orbital eccentricity. The
radial equation of motion can be written in the dimensionless form 
\be
{d^2 \xi \over dt^2} = 
{\angmom \over \xi^3} - {m(t) \over \xi^2} \, , 
\label{modeleq} 
\ee
where $\angmom$ is constant and where we have defined a 
dimensionless (total) mass 
\be
m(t) \equiv {M(t) \over M_{0\ast}} \,. 
\ee 
The dimensionless radial coordinate $\xi=r/a_0$ and the dimensionless
time variable is given in units of $\Omega^{-1}$ = 
$(a_0^3/G M_{0\ast})^{1/2}$.

The goal of this work is to find general solutions to the problem
where the dimensionless mass $m(t)$ monotonically decreases with
time. In the reduction of the standard two-body problem to an
analogous one-body problem, the equation of motion describes the orbit
of the reduced mass. In the version of the problem with mass loss
represented by equation (\ref{modeleq}), the motion is also that of a
reduced mass (e.g., Jeans 1924, MacMillan 1925, Hadjidemetriou 1963).
In this treatment, the mass loss functions (defined below) refer to
the dimensionless scaled mass $m(t)$. The model equation (\ref{modeleq}) 
is exact in the limit where the planet mass is small compared to the
stellar mass, i.e., $M_P/M_\ast \to 0$. For finite planetary masses,
the scaling between the two-body problem and the equivalent single
body problem changes quantities by ${\cal O}(M_P/M_\ast)$. In
applications of interest, however, our choice of mass loss functions
(and their uncertainties) provides the greatest degree of
approximation -- much larger than than the ${\cal O}(M_P/M_\ast)$
difference between the reduced problem and the full problem.

This paper thus focuses on the dimensionless problem of equation
(\ref{modeleq}).  The starting semimajor axis is unity (by definition)
and the initial conditions require that the starting radial coordinate
$\xi_0$ lies in the range $1-e \le \xi_0 \le 1+e$, where the starting
eccentricity $e$ is given by $\angmom = 1 - e^2$. Note that choosing
the value of $\xi_0$ is equivalent to specifying the starting phase of
the orbit (up to a sign).  The initial energy $\energy_0 = -1/2$ by 
definition and the initial (radial) velocity ${\dot \xi}_0$ is given by 
\be
{\dot \xi}_0^2 = {2\xi_0 - \angmom -\xi_0^2 \over \xi_0^2}= 
{[(1 + e) - \xi_0] [\xi_0 - (1-e)] \over \xi_0^2} \, . 
\ee
The initial velocity can be positive or negative, where the choice of
sign completes the specification of the starting phase of the orbit.

\subsection{Change of Variables} 
\label{sec:vchange} 

The equation of motion (\ref{modeleq}) is complicated because it
contains an arbitrary function, namely $m(t)$, that describes the mass
loss history. On the other hand, the independent variable (time) does
not appear explicitly. As a result, we may define a new effective 
``time'' variable $u$ through the expression 
\be
u \equiv {1 \over m} \, , 
\ee
where $m=m(t)$.  The generalized time variable $u$ starts at $u=1$ and
is monotonically increasing.  In terms of the variable $u$, the basic
equation of motion (\ref{modeleq}) takes the equivalent form
\be
{\dot u}^2 {d^2 \xi \over du^2} + {\ddot u} {d\xi\over du} 
= {\angmom \over \xi^3} - {1 \over u \xi^2} \, . 
\ee

Next we note that both standard lore and numerical solutions
(beginning with Jeans 1924) show that, in physical units, the product
$aM \approx$ {\sl constant}.  In terms of the current dimensionless
variables, this finding implies that the function 
\be
f \equiv {\xi \over u} = \xi m 
\ee
should vary over a limited range. We thus change the dependent
variable from $\xi$ to $f$ and write the equation of motion in the
form 
\be
{\dot u}^2 u^2 f^3 
\left[ \left( u^2 f^{\prime\prime} + 2u f^\prime \right) + 
{u {\ddot u} \over {\dot u}^2} \left( u f^\prime + f \right) \right] 
= \angmom - f \, , 
\label{fugeneral} 
\ee
where primes denote derivatives with respect to the variable $u$.
Keep in mind that equation (\ref{fugeneral}) is equivalent to the
original equation of motion (\ref{modeleq}), with a change in both the
independent and dependent variables.

The leading coefficient in equation (\ref{fugeneral}) represents an
important quantity in the problem: Note that the time scale for mass
loss is given by $u/{\dot u}$ and the orbital time scale is given by
$u^{2} f^{3/2}$ (this latter time scale is the inverse of the orbital
frequency, and is shorter than the orbital period by a factor of
$2\pi$). The ratio $\tratio$ of these two fundamental time scales is
given by 
\be 
\tratio^2 \equiv {\dot u}^2 u^2 f^{3} \, .
\label{timeratio} 
\ee
The leading coefficient in equation (\ref{fugeneral}) is thus
$\tratio^2$, the square of the ratio of the orbital time scale to the
time scale for mass loss.  For small values of $\tratio^2$, the mass
loss time is long compared to the orbit time, and the orbits are
expected to be nearly Keplerian; for larger $\tratio^2$, the star
loses a significant amount of mass per orbit and a Keplerian
description is no longer valid. For the former case, where mass loss
is slow compared to the orbit time, we can use the parameter $\tratio$
to order the terms in our analytic estimates. 

In addition to the coefficient $\tratio^2$, given by the ratio of time
scales, another important feature of equation (\ref{fugeneral}) is the
index $\undex$ appearing within the square brackets, where 
\be
\undex \equiv {u {\ddot u} \over {\dot u}^2 } \, . 
\label{defindex} 
\ee
The index $\undex$ encapsulates the time dependence of the mass loss.
This paper will focus on model equations with constant $\undex$ (such
models have a long history, from Jeans 1924 to Section
\ref{sec:mlossfun}).

{\sl The Orbital Energy:} For the chosen set of dimensionless
variables, the energy $\energy$ of the system takes the form 
\be
\energy = {1 \over 2} {\dot u}^2 \left( uf^\prime + f \right)^2 
+ {\angmom \over 2 u^2 f^2} - {1 \over u^2 f} \, . 
\label{energyfun} 
\ee
The energy has a starting value $\energy=-1/2$, by definition, and
increases as mass loss proceeds. If and when the energy becomes
positive, the planet is unbound. Although the energy expression
(\ref{energyfun}) appears somewhat complicated, the time dependence 
of the energy reduces to the simple form
\be
{d\energy \over du} = {1 \over u^3 f} \, . 
\label{energydq} 
\ee 
Note that the derivative of the energy is positive definite, so that
the energy always increases. Since the energy is negative and strictly
increasing, the semimajor axis of the orbit, when defined according to  
$a\propto|\energy|^{-1}$, is also monotonically increasing.  

\subsection{Mass Loss Functions} 
\label{sec:mlossfun} 

Next we want to specialize to the class of mass loss functions where
$\undex$ = {\sl constant}. The defining equation (\ref{defindex}) for
the mass loss index can be integrated to obtain the form 
\be
{\dot u} = \gamma u^\undex \, , 
\label{udotcon} 
\ee 
where $\gamma$ is a constant that defines the mass loss rate at the
beginning of the epoch (when $t=0$, $m=1$, and $u=1$). For a given 
(constant) value of the index $\undex$, the dimensionless mass loss 
rate has the form 
\be
{\dot m} = - \gamma m^{(2-\undex)} \, . 
\label{mform} 
\ee
In addition to simplifying the equation of motion, this form for the
mass loss function is motivated by stellar behavior, as discussed
below. The dimensionless parameter $\gamma$ is defined to be the ratio
of the initial orbital time scale to the initial mass loss time scale. 
Specifically, if we define $\tau=(M_{\ast}/{\dot M}_\ast)_0$, then 
$\gamma$ is given by 
\be
\gamma = {1 \over \tau} \left( {a_0^3 \over G M_{0\ast}} \right)^{1/2} 
\label{gammadef} 
\ee
$$
\approx 
1.6 \times 10^{-7} \left( {\tau \over 1 {\rm Myr} } \right)^{-1} 
\left( {a_0 \over 1 {\rm AU} } \right)^{3/2} 
\left( {M_{0\ast} \over 1 M_\odot } \right)^{-1/2} \,,
$$
where $M_{0\ast}$ is the stellar mass and $a_0$ is the semimajor axis 
at $t$ = 0. For typical orbits, $a_0$ = 1 -- 100 AU, so that we 
expect the parameter $\gamma$ to be small, often in the range 
$10^{-7}\lta\gamma\lta10^{-4}$. 

The mass loss rate of stars is often characterized by the physically 
motivated form 
\be
{\dot M} = - {\dot M}_C \left( {L_\ast \over L_\odot} \right) 
\left( {R_\ast \over R_\odot} \right) 
\left( {M_\ast \over M_\odot} \right)^{-1} \, , 
\label{physical} 
\ee   
where ${\dot M}_C$ is constant and depends on the phase of stellar
evolution under consideration (Kudritzki \& Reimers 1978, Hurley et
al. 2000). Since the radius and luminosity depend on stellar mass (for
a given metallicity), the physically motivated expression of equation
(\ref{physical}) can take the same power-law form as equation
(\ref{mform}), which corresponds to a constant mass loss index (see
equations [\ref{defindex}] and [\ref{udotcon}]).  

Using the scaling law (\ref{physical}), the power-law index appearing
in equation (\ref{mform}) can be positive or negative, depending on
how the stellar luminosity and radius vary with mass during the
different phases of mass loss (see Hurley et al. 2000 for a detailed
discussion). For example, if we consider main-sequence stars, the
stellar cores adjust quickly enough that the luminsoity obeys the
standard mass-luminosity relationship $L_\ast \sim M_\ast^p$ (where
the index $p\approx3$) and the mass-radius relationship $R_\ast \sim
M_\ast^q$ (where the index $q$ typically falls in the range
$1/2\le{q}\le1$). For main-sequence stars we thus obtain the scaling
law ${\dot m} \sim - m^{\alpha_m}$, where the index $\alpha_m=p+q-1$ is
predicted to lie in the range $2.5\le\alpha_m\le 3$; the corresponding
mass loss index lies in the range $-1\le\undex\le-1/2$.  Next we
consider stars on the first giant branch or the asymptotic giant
branch. In this phase of stellar evolution, mass loss occurs from an
extended stellar envelope, but the luminosity is produced deep within
the stellar core. As the star loses mass, the core and hence the
luminosity remains relatively constant, whereas the radius scales
approximately as $R_\ast \sim M_\ast^{-1/3}$ (Hurley et al. 2000).
For this case, one obtains the scaling law ${\dot m}\sim-m^{-4/3}$,
with a mass loss index $\undex$ = 10/3.  In general, for ${\dot m}
\sim - m^{\alpha_m}$, the mass loss index $\undex=2-\alpha_m$. As these
examples show, the mass loss index can take on a wide range of values 
$-1\le\undex\le4$. 

To fix ideas, we consider the time dependence for mass loss functions
that are often used. For a constant mass loss rate, the most common
assumption in the literature, the index $\undex$ = 2, and the mass 
evolution function has the form 
\be 
m(t) = 1 - \gamma t \qquad {\rm and} \qquad 
u(t) = {1 \over 1 - \gamma t} \, .
\label{mtconstant} 
\ee 
The value $\undex$ = 2 marks the boundary between models where the
mass loss rate accelerates with time ($\undex>2$) and those that
decelerate ($\undex<2$). For the case of exponential time dependence
of the stellar mass, the index $\undex$ = 1, and the mass loss
function has the form 
\be
m(t) = \exp[-\gamma t] \qquad {\rm and} \qquad 
u(t) = \exp[\gamma t] \,. 
\label{mtexpdecay} 
\ee
The value $\undex$ = 1 marks the boundary between models where the
system reaches zero stellar mass in a finite time ($\undex>1$), and
those for which the mass $m\to0$ only in the limit $u\to\infty$. 
For the case with index $\undex$ = 0, which represents an important
test case, the mass evolution function becomes
\be
m(t) = {1 \over 1 + \gamma t} \qquad {\rm and} \qquad 
u(t) = {1 + \gamma t} \,. 
\label{mtzero} 
\ee
For $\undex$ = 0, analytic solutions are available (see Section
\ref{sec:bzero}), which inform approximate treatments for more general
values of the index $\undex$.  Finally, the case where $\undex=-1$ 
plays a defining role (see Section \ref{sec:bminusone}) and 
corresponds to the forms
\be
m(t) = (1 + 2 \gamma t)^{-1/2} 
\qquad {\rm and} \qquad 
u(t) = (1 + 2 \gamma t)^{1/2} \,. 
\label{mttransition} 
\ee
The value $\undex=-1$ marks the boundary between models where the
planet becomes unbound at finite stellar mass ($\undex>-1$) and those
for which the planet becomes unbound only in the limit $m\to0$ or
$u\to\infty$ ($\undex<-1$). In general, for constant $\undex \ne 1$, 
the time dependence of the mass takes the form 
\be
m(t) = {1 \over u(t)} = 
\left[ 1 - (\undex - 1) \gamma t \right]^{1/(\undex-1)} \, .
\label{mtgeneral} 
\ee 
The particular case $\undex=1$ results in the decaying
exponential law of equation (\ref{mtexpdecay}). 

\subsection{Equation of Motion with Constant Index $\undex$} 
\label{sec:equamotion} 

For constant values of the mass loss index $\undex$, the 
equation of motion reduces to the form
\be
\tratio^2 \left[ u^2 f^{\prime\prime} + (2 + \undex) u f^\prime + 
\undex f \right] = \angmom - f \, , 
\label{fuconst} 
\ee
where the ratio of time scales $\tratio$ is given by 
\be
\tratio^2 = \gamma^2 u^{2\undex + 2} f^3 \, . 
\label{tratiocon} 
\ee 

\noindent 
By writing the equation of motion in the form (\ref{fuconst}), we
immediately see several key features of the solutions: 

When the parameter $\tratio \ll 1$, the left-hand side of equation
(\ref{fuconst}) is negligible, and the equation of motion reduces to
the approximate form $f \approx \angmom$ = {\sl constant}. This
equality is only approximate, because the function $f$ also executes
small oscillations about its mean value as the orbit traces through
its nearly elliptical path (see below). Nonetheless, this behavior is
often seen in numerical studies of planetary systems with stellar mass
loss (e.g., see Debes \& Sigurdsson 2002).  Orbital evolution with
$\tratio\ll1$ is often called the ``adiabatic regime''. We note that
this terminology is misleading, however, because ``adiabatic'' refers
to evolution of a thermodynamic system at constant energy (heat),
whereas the systems in question steadily gain energy through stellar
mass loss (the gravitational potential becomes less negative).

When the parameter $\tratio \gg 1$, the left-hand side of equation
(\ref{fuconst}) dominates, and the solutions for $f(u)$ take the form 
of power-laws with negative indices. In this regime, the equation of 
motion approaches the form 
\be
u^2 f^{\prime\prime} + (2 + \undex) u f^\prime 
+ \undex f = 0 \, , 
\ee 
so that the function $f(u)$ has power-law solutions with indices
$p$ given by the quadratic equation 
\be
(p + 1) (p + \undex) = 0 \, . 
\ee
The general form for the solution $f(u)$ in this regime is thus  
\be
f(u) = {A \over u} + {B \over u^\undex} \, . 
\label{fastsol} 
\ee
After the solutions enter this power-law regime, the energy can
quickly grow and the planet can become unbound. To illustrate this
behavior, consider the differential equation (\ref{energydq}) for the
orbital energy. We first consider the regime where $\tratio \ll 1$ and
the function $f$ is nearly constant.  For the benchmark case $f$ = 1, 
the equation can be integrated to obtain
\be
\energy = - {1 \over 2 u^2} \, . 
\ee
As long as $f\approx1$, the energy remains negative and the planet
remains bound, except in the limit $u \to \infty$. Now let $\tratio>1$
so that the solutions enter into the power-law regime. If we let the 
solution have the form $f=A/u$, for $u>u_c$, the differential equation 
for the energy can be integrated to obtain 
\be
\energy = \energy_c + {1 \over A u_c} 
\left( 1 - {u_c \over u} \right) \, , 
\ee
where the subscript $c$ denotes the reference point where the
solutions enters into the power-law regime. Since $A u_c \sim u_c^2$
and $|\energy_c| \sim u_c^2/2$, the energy quickly becomes positive
once the power-law regime is reached. This argument indicates that the
planet becomes unbound when the time scale ratio $\tratio$ is of order
unity (as seen in previous studies).  The subsequent subsections
provide further verification of this finding. 

In order for the solution to make the transition from $f\approx$ 
{\sl constant} to the power-law solutions that cause the orbits to
become unbound, the ratio of time scales $\tratio$ must grow with
time.  However, growth requires that $\undex > -1$ (see equation
[\ref{tratiocon}]). We can understand this requirement as follows: 
The orbital time scale $P$ varies with $u$ (and hence time) according
to $P \sim u^2 f^{3/2}$. Since $f$ is nearly constant, this relation
simplifies to the form $P \sim u^2$. The time scale for mass loss
$\tau$ is given by $\tau = u/{\dot u}$, which has the form $\tau\sim
u^{1 - \undex}$ from equation (\ref{udotcon}). As a result, when
$\undex=-1$, the orbit time has the same dependence on stellar mass as
the mass loss time scale, so that the ratio $\tratio$ is nearly
constant as the star loses mass. For $\undex<-1$, the ratio $\tratio$
of time scales decreases with time, and the system grows ``more
stable''.

\section{Results for Single Planet Systems with Stellar Mass Loss} 
\label{sec:results} 

This section presents the main results of this paper for single planet
systems with a central star that loses mass.  First, we consider mass
loss index $\undex=-1$, which marks the critical value such that
systems with $\undex>-1$ become unbound at finite values of the
stellar mass, whereas systems with $\undex<-1$ only become unbound in
the limit $m\to0$. Next we consider mass loss index $\undex$ = 0; in
this case, the solutions can be found analytically, and these results
guide an approximate analytic treatment of the general case, which is
addressed next. We also consider the limiting case where stellar mass
loss takes place rapidly.

\subsection{The Transition Case} 
\label{sec:bminusone} 

Here we consider systems where the mass loss index $\undex=-1$, which
corresponds to the transition value between cases where the ratio
$\tratio$ of time scales grows with time ($\undex>-1$) and those where
the ratio decreases with time ($\undex<-1$). In this regime, the equation 
of motion reduces to the form 
\be
\gamma^2 \left[ u^2 f^{\prime\prime} + u f^\prime - f \right] 
= {\angmom \over f^3}  - {1 \over f^2} \, . 
\ee 
The equation of motion can be simplified further by making the 
change of (independent) variable 
\be
w \equiv \log u \, , 
\ee
so that the equation of motion becomes 
\be
\gamma^2 \left[ {d^2 f \over dw^2} - f \right] 
= {\angmom \over f^3}  - {1 \over f^2} \, . 
\ee 
This version of the equation of motion (first considered by Jeans
1924) contains no explicit dependence on the independent variable $w$,
so that the equation can be integrated to obtain the expression 
\be
\gamma^2 \left( {df \over dw} \right)^2 = \gamma^2 f^2 + 
{2 \over f} - {\angmom \over f^2} - E \, , 
\ee
where $E$ is a constant that plays the role of energy. Note that we
have chosen the sign such that $E>0$ and that $E=1$ for initially
circular orbits. In order for the function $f(w)$ to have oscillatory
solutions, the fourth order polynomial  
\be
p(f) = \gamma^2 f^4 - E f^2 + 2f - \angmom 
\ee
must be positive between two positive values of $f$. In order for this
requirement to be met, the parameters must satisfy the inequalities
\be
\angmom E \le {9 \over 8 } \qquad {\rm and} \qquad 
\gamma \le (E/3)^{3/4} \, . 
\label{gammax} 
\ee
The first inequality is always satisfied for the cases of interest.
The second inequality in equation (\ref{gammax}) determines the
maximum value of the mass loss parameter $\gamma$ for which
oscillatory solutions occur.

\subsection{Systems with Vanishing Mass Loss Index} 
\label{sec:bzero} 

In the particular case where $\undex$ = 0, the equation of motion can 
be simplified. In particular, the first integral can be taken analytically 
to obtain the form 
\be
\gamma^2 \left(u^2 f^\prime\right)^2 = 
- {\angmom \over f^2} + {2 \over f} - E \, , 
\label{firstint} 
\ee 
where $E$ is a constant of integration. The parameter $E$ plays the
role of energy for the orbit problem where the function $f(u)$ plays
the role of the radial coordinate.  Although $E$ is constant, the
energy $\energy$ of the physical orbit (where $\xi$ is the radial
coordinate) increases with time. Notice also that we have adopted a
sign convention so that $E>0$. The value of $E$ depends on the initial
configuration. For the particular case where the orbit starts at
periastron, for example, the initial speed ${\dot \xi}$ = 0 and the
energy constant has the value 
\be
E = 1 - \gamma^2 (1 - e)^2 \, , 
\label{edefzero} 
\ee
where the eccentricity $e$ = $(1-\angmom)^{1/2}$. In general, the 
initial value $f_0 = \xi_0$ can lie anywhere in the range 
$1-e \le f_0 \le 1+e$, and the energy constant has the general form 
\be
E = 1 - \gamma^2 f_0^2 \pm 2 \gamma 
\left[ 2 f_0 - \angmom - f_0^2 \right]^{1/2} \,, 
\label{edefine} 
\ee
where the choice of sign is determined by whether the planet is 
initially moving outward ($+$) or inward ($-$). With the energy 
constant $E$ specified, the turning points for the function $f$ 
are found to be 
\be
f_{1,2} = {1 \pm \left[ 1 - \angmom E \right]^{1/2} \over E} \, . 
\label{turningpts} 
\ee
If we consider the function $f$ to play the role of the radial 
coordinate, then equation (\ref{turningpts}) defines analogs of 
the semimajor axis $a_\ast$ and eccentricity $e_\ast$, 
which are given by 
\be 
a_\ast = {1 \over E} \qquad {\rm and} \qquad 
e_\ast = \left[ 1 - \angmom E \right]^{1/2} \, .
\label{zelements} 
\ee
For a given starting value $f_0$, the effective eccentricity
is given by the expression 
\be
e_\ast^2 = e^2 + \gamma^2 (1 - e^2) f_0^2 \mp 2 \gamma
(1 - e^2) \left( 2f_0 - \angmom - f_0^2 \right)^{1/2} .
\label{estar} 
\ee
Note that the effective eccentricity $e_\ast$ of the function $f$
is larger than the initial eccentricity $e$ of the original orbit 
(before the epoch of mass loss). In particular, for a starting circular 
orbit $e=0$, the effective eccentricity $e_\ast = \gamma \ne 0$. 
The integrated equation of motion (\ref{firstint}) can be 
separated and written in the form 
\be
{f df \over \left[ (f - f_1) (f_2 - f) \right]^{1/2} } = 
{E^{1/2} \over \gamma} \, {du \over u^2} \, . 
\ee
If we integrate this equation from one turning point to the other, 
the change in mass of the system the same for every cycle, i.e.,  
\be
\Delta m = {\gamma \pi \over E^{3/2} } \, , 
\ee
where $E$ is given by equation (\ref{edefine}). Following standard 
procedures, we can find the solution for the orbit shape, which can 
be written in the form 
\be
{\angmom \over f} = {\angmom \, u \over \xi} = 1 + e_\ast \cos\theta \, . 
\ee
The orbit equation thus takes the usual form, except that the original
eccentricity $e$ is replaced with the effective eccentricity $e_\ast$
and the effective ``radius'' variable ($f$) scales with the mass/time 
variable $u=1/m$. 

For this mass loss function (with $\undex$ = 0), we can find a simple
relationship between the value of the time scale ratio $\tratio$ and
the value of $f$ when the planet becomes unbound. To obtain this
result, we insert the first integral from equation (\ref{firstint})
into the general expression (\ref{energyfun}) for the energy $\energy$
of the orbit and set $\energy$ = 0. After eliminating the derivative
$f^\prime$, we can solve for the time scale ratio $\tratio_f$ as a
function of the final value of $f$.  When the planet becomes unbound, 
the time scale ratio is thus given by
\be
\tratio_f = { \left( 2f - \angmom \right)^{1/2} \pm 
\left(2f - \angmom - f^2 \right)^{1/2} \over f^{1/2}} \,. 
\label{tratvf} 
\ee
Here, $f$ is evaluated when the planet becomes unbound. In this
case, however, the value of $f$ is constrained to lie in the range
$f_1 \le f \le f_2$, where the turning points are given by equation
(\ref{turningpts}). For small $\gamma$, the orbit oscillates back and
forth between the turning points many times before the planet becomes
unbound. The final value of $f$ is thus an extremely sensitive
function of the starting orbital phase. This extreme sensitivity is
not due to chaos, and can be calculated if one knows the exact orbital 
phase at the start. In practice, however, the final value of $f$ can be 
anywhere in the range $f_1 \le f \le f_2$. 

\begin{figure} 
\centerline{\epsscale{0.90} \plotone{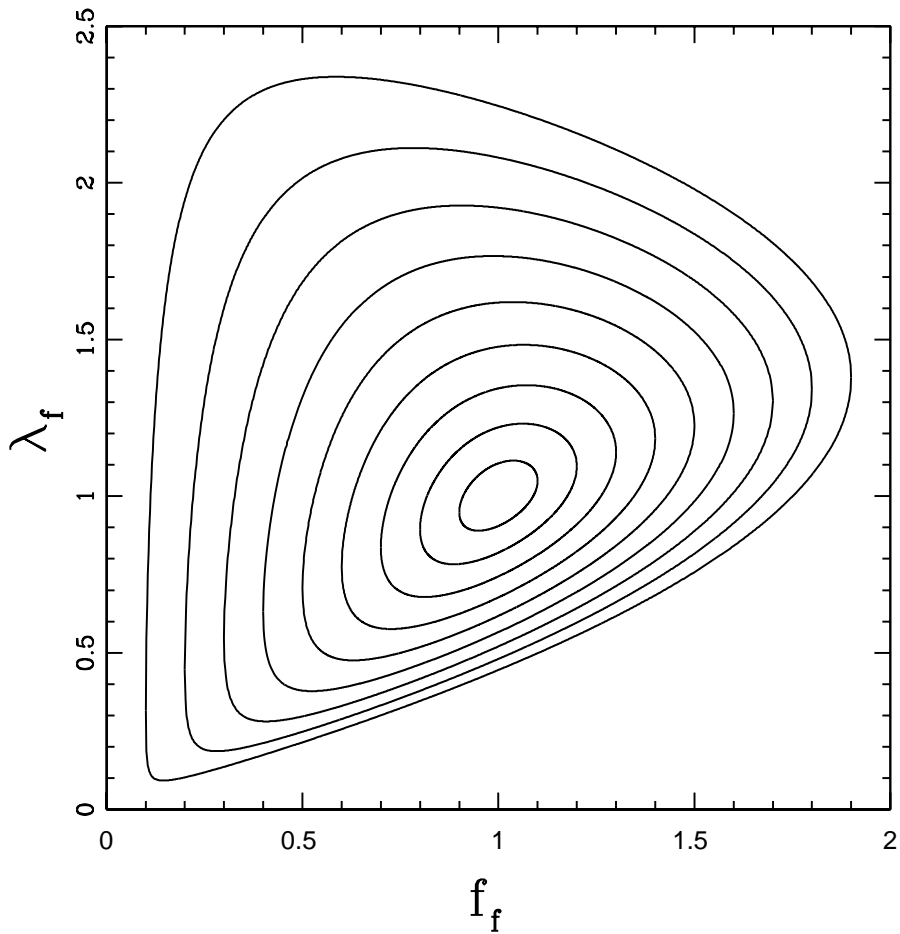}} 
\figcaption{Time scale ratio $\tratio_f$ as a function of the final 
value of $f=f_f$ for planetary systems with $\undex$ = 0. For a given
angular momentum $\angmom$, specified by the starting eccentricity,
the allowed values of $\tratio_f$ form closed curves in the plane as
shown. Curves are shown for a range of starting eccentricity, from 
$e$ = 0.1 (inner curve) to $e$ = 0.9 (outer curve). } 
\label{fig:bzeroplane} 
\end{figure}

Figure \ref{fig:bzeroplane} shows the final values of the time scale
ratio $\tratio$ as a function of the final value of $f=\xi/u=\xi{m}$.
Curves are shown for nine values of the starting eccentricity, where
$e$ = 0.1, 0.2, $\dots$ 0.9. The innermost (outermost) closed curve in
the figure corresponds to the smallest (largest) eccentricity.  Each
value of $f$ corresponds to two possible values of the time scale
ratio $\tratio$, one for orbits that are increasing in $f$ and one for
orbits that are decreasing in $f$ at the time when the planet becomes
unbound.  These two values of $\tratio$ (for a given $f$) correspond
to the two branches of the solution given by equation (\ref{tratvf}).

\begin{figure} 
\centerline{\epsscale{0.90} \plotone{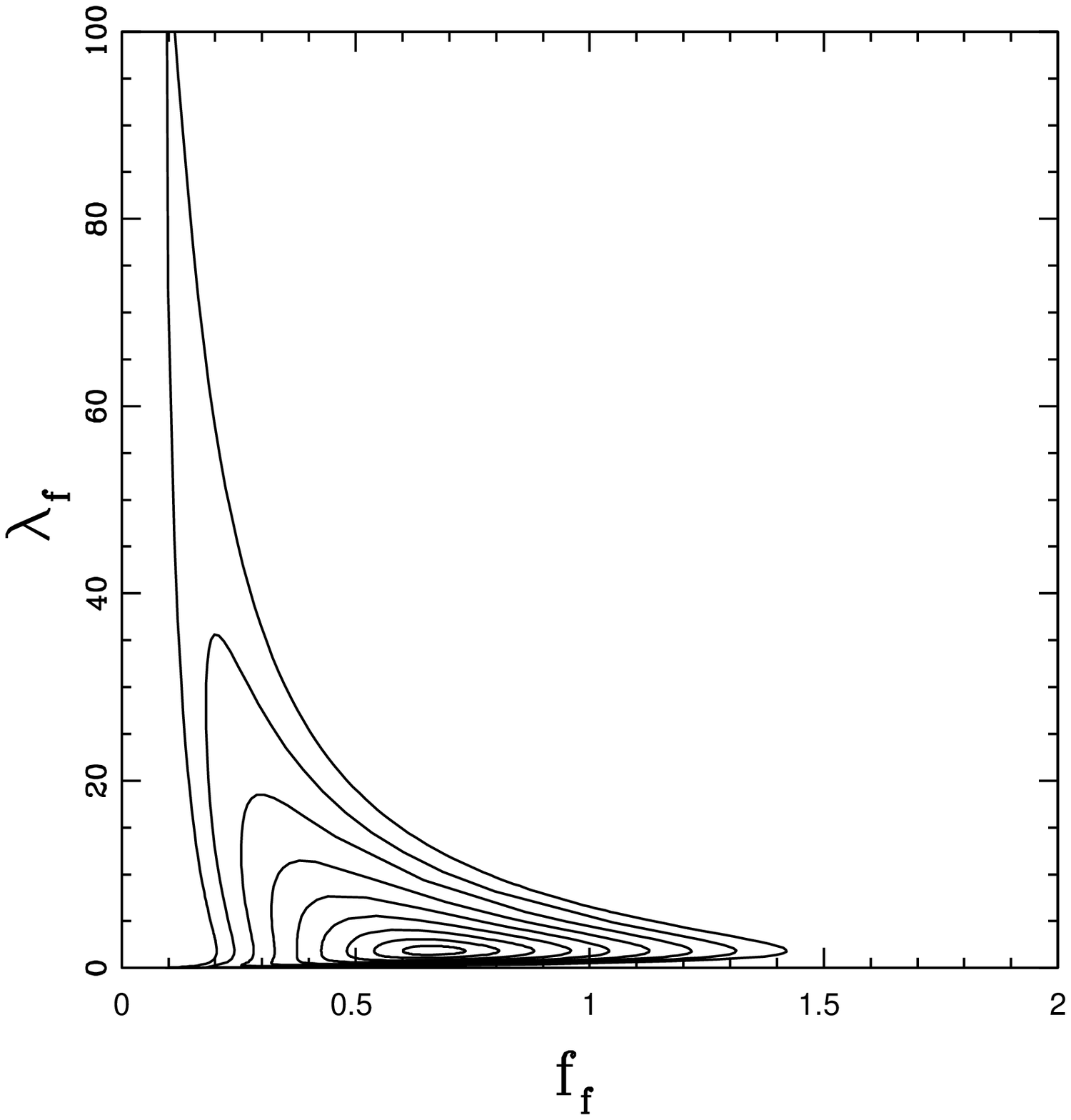} }  
\figcaption{Time scale ratio $\tratio_f$ as a function of the final 
value of $f=f_f$ for planetary systems with $\undex$ = 2 (constant
mass loss rates). For a given angular momentum $\angmom$, specified 
by the starting eccentricity, the allowed values of $\tratio_f$ form
closed curves in the plane as shown. Curves are shown for a range of
starting eccentricity, from $e$ = 0.1 (inner curve) to $e$ = 0.9
(outer curve).  Compare with Figure \ref{fig:bzeroplane} (and note 
the change of scale). }
\label{fig:btwoplane} 
\end{figure}

For comparison, Figure \ref{fig:btwoplane} shows the same plane of
parameters for the final values of the time scale ratio $\tratio_f$
and $f_f$ for planetary systems with constant mass loss rate (where
$\undex=2$). These results were obtained through numerical integration
of equation (\ref{fuconst}).  Note that the range of allowed values
for the time scale ratio $\tratio_f$ is much larger than for the case
with $\undex$ = 0, whereas the range of final values $f_f$ is somewhat
smaller.

One can also show that for circular orbits ($\angmom=1$ and $e=0$),
the value of the time scale ratio $\tratio$ = 1 when the planet
becomes unbound. For circular orbits in the limit $\gamma\to0$, the
turning points of the orbit appraoch $f_{1,2}$ = 1. Using $f=1$ and
$\angmom=1$ in equation (\ref{tratvf}), we find $\tratio=1$. 

\subsection{Limit of Rapid Mass Loss} 
\label{sec:rapidloss} 

If we now consider the case where the mass loss is rapid, so that the
equation of motion has solution (\ref{fastsol}) throughout the
evolution, we can fix the constants $A$ and $B$ by applying the
initial conditions. Since $f = \xi/u$ and $u=1$ at the start of the
epoch, $f(1)$ = $\xi_1$, where $\xi_1$ is the starting value of the
orbital radius. By definition, the semimajor axis is unity, and the
starting orbital eccentricity is given by $e^2=1-\angmom$. The starting 
radius thus lies in the range 
\be
1 - \sqrt{1 - \angmom} \le \xi_1 \le 1 + \sqrt{1 - \angmom} \, , 
\label{range} 
\ee
which is equivalent to $1-e\le\xi_1\le1+e$.
The derivative $f^\prime = df/du$ is given by 
\be
f^\prime = 
- {\xi \over u^2} + {\xi^\prime \over u} = 
- {\xi \over u^2} + {1\over\gamma u^{\undex+1}} 
{d\xi \over dt} \, . 
\ee
In the regime of interest where $\gamma \gg 1$, the second term is
small compared to the first. In this limit, $f^\prime(1)=-\xi_1$,
where $\xi_1$ lies in the range indicated by equation (\ref{range}).
The constants $A$ and $B$ are thus determined to be $B$ = 0 and 
$A=\xi_1$, so that the solution has the simple form 
\be
f(u) = {\xi_1 \over u} \, . 
\label{solution} 
\ee
The energy of the orbit, by definition, starts at
$\energy(u=1)=\energy_1=-1/2$, and the energy obeys the differential
equation (\ref{energydq}).  Combining the solution of equation
(\ref{solution}) with the differential equation (\ref{energydq}) for
energy, we can integrate to find the energy as a function of $u$
(equivalent to time or mass),
\be
\energy = - {1 \over 2} + {1 \over \xi_1} 
\left( 1 - {1 \over u} \right) \, . 
\ee
We can then read off the value of $u_f$, and hence the mass $m_f$, 
where the energy becomes positive and the planet becomes unbound, i.e., 
\be
m_f = {1 \over u_f} = 1 - {\xi_1 \over 2} \, . 
\label{masslimit} 
\ee
Note that this critical value of the mass depends on the orbital phase
of the planet within its orbit, i.e., the result depends on $\xi_1$
rather than the starting semimajor axis, which is unity (notice also
that this condition is equivalent to that given by equations [46--48]
in Veras et al. 2011).  For circular orbits, we must have $\xi_1$ = 1, 
so that planets become unbound when the stellar mass decreases by one 
half (as expected). 

For cases where the mass loss is rapid, but the planet remains bound,
we can find the orbital properties for the post-mass-loss system.
Consider the limiting case where the star has initial mass $m$ = 1,
and loses a fraction of its mass instantly so that it has a final mass
$m_\infty$, i.e., 
\be
m(t) = m_\infty + (1 - m_\infty) H(-t) \, ,  
\ee
where $H$ is the Heaviside step function. The mass loss thus occurs
instantaneously at $t$ = 0. For $t<0$, the solutions to the orbit
equation (\ref{modeleq}) have the usual form, 
\be
{\dot \xi}^2 = {2 \over \xi} - {\angmom \over \xi^2} - 1 = 
{ (\xi - \xi_1) (\xi_2 - \xi) \over \xi^2} \, ,  
\ee
where $\xi_{2,1} = 1 \pm e$ and $\angmom = 1 - e^2$. After mass loss
has taken place, the new (dimensionless) stellar mass is $m_\infty$, 
and the first integral of the equation of motion can be written in 
the form 
\be
{\dot \xi}^2 = {2m_\infty\over\xi} - {\angmom\over\xi^2} - 1 
+ {2(1-m_\infty)\over\xi_0} \, , 
\ee
where the final constant term takes into account the change in
(dimensionless) energy at the moment of mass loss. The radial position
at $t$ = 0 is $\xi_0$; since the planet is initially in a bound
elliptical orbit, the radial coordinate must lie in the range 
$1-e \le \xi_0 \le 1+e$. The energy $\energy_f$ of the new orbit 
is thus given by 
\be
2 \energy_f = - 1 + {2(1-m_\infty)\over\xi_0} \, . 
\label{newenergy} 
\ee  
The energy $\energy_f$ is negative, and the orbit is bound, provided
that the remaining mass $m_\infty > 1 - \xi_0/2$. This condition is
thus consistent with equation (\ref{masslimit}), which defines the
mass scale at which orbits become unbound in the limit of rapid mass
loss. In terms of the energy $\energy_f$, the turning points of the 
new orbit take the form 
\be
\xi_{\pm} = {m_\infty \pm \left[ m_\infty^2 - 2 |\energy_f| 
\angmom \right]^{1/2} \over 2 |\energy_f|} \, . 
\ee
We can then read off the orbital elements for the new 
(post-mass-loss) orbit, i.e., 
\be
a_f = {m_\infty \over 2 |\energy_f| } 
\qquad {\rm and} \qquad 
e_f = \sqrt{ 1 - 2 |\energy_f| \angmom / m_\infty^2 } \, , 
\label{newelements} 
\ee
where the new orbital energy $\energy_f$ is given by equation 
(\ref{newenergy}). 

\subsection{Systems with General Mass Loss Indices} 
\label{sec:generalind} 

In order to address the general case, we first change variables
according to the ansatz 
\be 
x = u^\alpha \qquad {\rm where} \qquad 
\alpha = \undex + 1 \, . 
\ee
After substitution, the equation of motion becomes 
\be
\gamma^2 \alpha^2 x^2 \left[ x^2 f_{x x} + 2 x f_x \right] 
+ \gamma^2 \undex x^2 f = {\angmom \over f^3} - {1 \over f^2} \, ,
\label{fxsimple}
\ee 
where the subscripts denote derivatives with respect to the new
variable $x$. The ratio of time scales is now given by 
\be
\tratio^2 = \gamma^2 x^2 f^3 \, .
\ee
We can integrate the differential equation (\ref{fxsimple}) to 
obtain the implicit form 
\be
\gamma^2 \alpha^2 \left[ x^2 f_x \right]^2 + 
2 \gamma^2 \undex \int_1^x x^2 f f_x dx 
=  - {\angmom \over f^2} + {2 \over f} - E \, ,
\label{fxint}
\ee 
where $E>0$ has the same meaning as before. 
To move forward, we define the integral quantity 
\be
J \equiv  2 \gamma^2 \undex \int_1^x x^2 f f_x dx \,,
\ee
so that 
\be
\gamma^2 \alpha^2 \left[ x^2 f_x \right]^2 + J 
=  {2f - \angmom - E f^2 \over f^2} \, . 
\label{fxfive}
\ee 
Note that $J={\cal O}(\tratio^2)$, which means that $J$ will 
be negligible for most of the mass loss epoch (see Appendix 
\ref{sec:jbound}). The energy of the system can be written 
in the form  
\be
u^2 \energy = {1 \over 2} \gamma^2 x^2 (\alpha x f_x + f)^2 + 
{\angmom \over 2 f^2} - {1 \over f} \, . 
\label{genenergy} 
\ee
At the start of the evolution (where $u$ = 1 and $x$ = 1), 
the energy $\energy = -1/2$ by definition. Using this specification, 
we can find the value of the integration constant $E$, which takes 
the form 
\be
E = 1 - \gamma^2 f_0^2 \pm 2\gamma 
\left( 2f_0 - \angmom - f_0^2 \right)^{1/2} \, , 
\ee
where $f_0(=\xi_0)$ is the starting value of the function 
(radial variable). 

At an arbitrary time during the epoch of mass loss, 
we can write the derivative $f_x = df/dx$ in the form  
\be
\gamma \alpha \left[ x^2 f_x \right] =  \pm 
{1 \over f} \left( 2f - \angmom - E f^2 - J f^2 \right)^{1/2} \, . 
\label{fxsix}
\ee 
Since $J={\cal O}(\tratio^2)$, the condition $|J| \ll E$ holds for 
most times. As a result, working to leading order, we can set $J$ = 0
in equation (\ref{fxsix}) and recover analogs to the orbital solutions
found earlier in Section \ref{sec:bzero} (for a related result, see
Radzievskii \& Gel'Fgat 1957; for an alternate approach, see Rahoma et
al. 2009). The only difference is that the dependent (time-like)
variable $u$ is replaced with $x = u^{\beta+1}$.  As a result, the
turning points for the function $f(x)$ will be given by equation
(\ref{turningpts}) and the orbital elements for $f(x)$ are given by
equation (\ref{zelements}).

\begin{figure} 
\centerline{\epsscale{0.90} \plotone{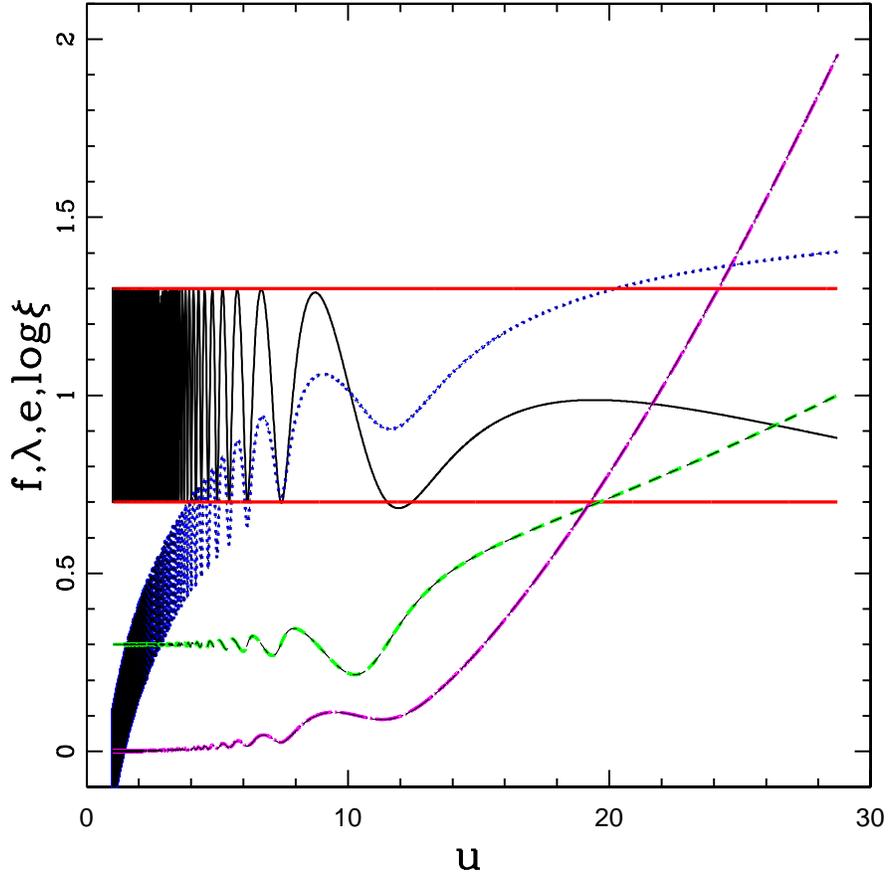} } 
\figcaption{Evolution of the orbit during the epoch of stellar mass 
loss. In this example, the mass loss function has index $\undex$ = 2,
corresponding to a constant mass loss rate. The other parameters are
$\gamma$ = $10^{-4}$, $e$ = 0.3, and $f_0$ = 1 (initially going inward). 
The black curve shows the function $f(u)=\xi/u$; the red horizontal 
lines mark the analytically determined turning points of the function. 
The blue dotted curve shows the evolution of the radial coordinate 
$\xi$ (plotted here as $\log_{10}[\xi]$). The magenta dot-dashed curve 
shows the evolution of the time scale ratio $\tratio$. Finally, the 
green dashed curve shows the eccentricity $e$ of the orbit. }  
\label{fig:limits}
\end{figure}

The basic behavior of the orbit is illustrated by Figure
\ref{fig:limits}.  The function $f$, plotted here versus $u$ as the
solid black curve, oscillates between the turning points (marked by
the red horizontal lines) given by equation (\ref{turningpts}).  The
radial coordinate (here $\log \xi$ is plotted as the dotted blue
curve) oscillates also, but grows steadily. The eccentricity of the
orbit (green dashed curve) also oscillates, but grows with time.
Finally, the time scale ratio $\tratio$ (magenta dot-dashed curve)
also oscillates and grows with time. The simple oscillatory behavior
for $f(u)$ ceases near the point where the time scale ratio $\tratio$
becomes of order unity. Note that the function $f$ falls outside the
boundary marked by the turning points near $u=12$ in the Figure.

Next we consider the time evolution of the energy of the orbit. 
After some rearrangement, the energy from equation (\ref{genenergy})
can be rewritten in the form
\be
2 u^2 \energy = - E - J 
\label{enerfull}
\ee
$$
\pm 2 \gamma x \left( 2f - \angmom - E f^2 - J f^2 \right)^{1/2} 
+ \gamma^2 x^2 f^2  \, . 
$$ 
Since $J$ = ${\cal O}(\tratio^2)$, it is often convenient to work 
in the limit $J\to0$ where the energy becomes 
\be
2 u^2 \energy = - E 
\pm 2 \gamma x \left( 2f - \angmom - E f^2 \right)^{1/2} 
+ \gamma^2 x^2 f^2  + {\cal O} \left( J \right) \, . 
\label{jzenergy}
\ee 
This form for the energy shows why the product $am$ of the effective
semimajor axis and the mass is slowly varying: In dimensionless units,
the energy $\energy = - m/2a$ so that 
\be
\left( am \right)^{-1} = - 2 \energy u^2 = E + 
{\cal O} \left( \tratio \right) \, . 
\ee 
The product $am$ is thus nearly constant as long as the time scale
ratio $\tratio$ is small, and the departure is of order $\tratio$.
When $\tratio$ is small, the orbit cycles through many turning points
before the mass changes substantially, so that the average of the
above equation becomes
\be
\langle \left( am \right)^{-1} \rangle = 
\langle - 2 \energy u^2 \rangle = E + 
{\cal O} \left( \tratio^2 \right) \, , 
\label{amconstant} 
\ee 
so that the average of $am$ is constant to second order. 

{\sl Number of Cycles:} If we ignore $J$ for now and integrate 
equation (\ref{fxint}) over one cycle, we obtain 
\be
{\gamma \alpha \over E^{1/2}} 
\int_1^2 {f df \over (f - f_1)^{1/2} (f_2 - f)^{1/2}} = 
\int_1^2 {dx \over x^2} = {1 \over x_1} - {1 \over x_2} \, . 
\label{cycle} 
\ee 
The integral on the LHS gives us $\pi/E$. 
If we integrate over $N$ cycles we obtain the expression 
\be
\gamma \alpha N \pi = E^{3/2} \left[ 1 - m_N^\alpha \right] \, , 
\ee
where we assume that $m=1$ at the start. The total number 
of possible cycles occurs when $m_N \to 0$, so that 
\be
N_T = {E^{3/2} \over \pi \gamma \alpha} = 
{E^{3/2} \over \pi \gamma (\undex + 1) } \,.
\label{numcycles} 
\ee  
Since $E \sim 1$ and $\pi(\beta+1)\sim10$, whereas $\gamma \ll 1$, we
expect the number of cycles $N_T \sim 1/(10\gamma)$ to often be large.

{\sl The Last Cycle:} The above analysis (if we continue to work in
the regime $J\ll1$) suggests that the last cycle occurs when the right
hand side of equation (\ref{cycle}) is no longer large enough to
balance the left hand side, which is the same for each cycle. As a
result, a minimum mass must be left in the star in order for the orbit
to complete a cycle (in the function $f$). This condition can be
written in the form 
\be
m_c = \left( {\pi \gamma \alpha \over E^{3/2}} \right)^{1/\alpha} \,,
\ee
which holds for $\alpha=1+\undex>0$. At the point when the mass falls
below this threshold, the time scale ratio is given by $\tratio$ =
$(Ef)^{3/2}/(\pi\alpha)$.  Since $Ef\sim1$ and $\pi\alpha\sim3-10$, the
system crosses the threshold so that $f$ cannot complete a cycle just
before the time scale ratio $\tratio$ reaches unity.

{\sl Final States:} If we set the energy equal to zero and replace the
variable $x$ with the time scale ratio $\tratio_f$ (evaluated at the
moment that planet becomes unbound), we find the condition 
\be
(2f - \angmom)^{1/2} = \tratio_f f^{1/2} \pm 
\left( 2f - \angmom - E f^2 - J f^2 \right)^{1/2} \,, 
\ee
which can then be written in the form 
\be
\tratio_f = {(2f - \angmom)^{1/2} \pm 
\left( 2f - \angmom - E f^2 - J f^2 \right)^{1/2} 
\over f^{1/2} } \,. 
\ee
This expression is thus a generalization of that obtained for the
special case with $\undex$ = 0 (see equation [\ref{tratvf}] and Figure
\ref{fig:bzeroplane}).  The differences are that we have included the
extra term $J$ and that the result is written in terms of the variable
$x$ instead of $u$.

\begin{figure} 
\centerline{\epsscale{0.90} \plotone{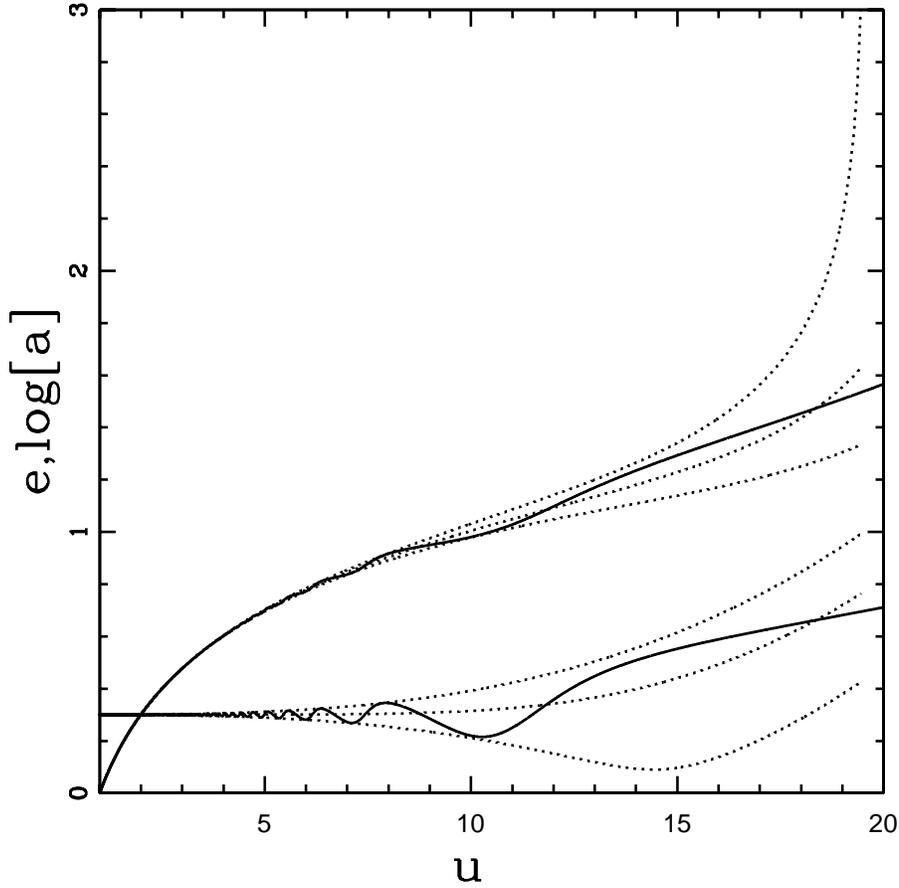} } 
\figcaption{Evolution of orbital elements during the epoch of stellar 
mass loss. In this example, the mass loss function has index $\undex$
= 2. The other parameters are $\gamma=10^{-4}$, $e=0.3$, and $f_0=1$ 
(initially going inward).  The solid curves show the semimajor axis 
($\log a$; top) and orbital eccentricity ($e$; bottom), calculated 
from numerical integration of the equation of motion. For each
orbital element, the three dotted curves show the average value, 
the upper limit, and the lower limit, as calculated from the analytic
expressions given by equations (\ref{enermean} -- \ref{efinal}). }
\label{fig:elements}
\end{figure}

The orbital eccentricity, calculated the usual way, oscillates with
time with an increasing amplitude of oscillation (e.g., see Figure
\ref{fig:elements}). As shown here, however, the function $f$ executes
nearly Keplerian behavior, with nearly constant turning points, where
this statement is exact in the limit $J\to0$. The oscillation of
eccentricity, although technically correct, is misleading. The turning
points of the orbit (in the original radial variable $\xi$) are
strictly increasing functions of time. The oscillation in $e$ arises
because the orbits are not ellipses, and, in part, because the period
of the orbits in $\xi$ are not the same as the period of the orbits in
$f$. As a result, the oscillations in the calculated eccentricity do
not imply that the near-elliptical shape of the orbit is varying
between states of greater and lesser elongation. Instead, these 
oscillations imply that if the mass loss stops and the orbit once
again becomes an ellipse, the value of the final eccentricity of that
ellipse oscillates with the ending time of the mass loss epoch. 

{\sl Orbital Elements during and after the Mass Loss Epoch:} Next we
consider the case where the planet remains bound after the epoch of
stellar mass loss. In this case, we want to estimate the orbital
elements of the planet. We are mostly interested in specifying the
orbital elements at the end of the mass loss epoch, but we can also
evaluate them at any time while the star continues to lose mass.
Suppose that the orbit passes through $N$ turning points of the
function $f$ during the mass loss epoch. The orbit will then complete
a partial cycle so that the final value of $f$ lies between the
turning points, $f_1\le{f}\le{f_2}$. In the ideal case, where we have
complete information describing both the starting orbital elements and
the final value of the stellar mass, and where the mass loss function
is is described exactly by a model with constant index $\undex$, we
can calculate the final value $f_f$. In practice, we will often have
incomplete information: The number of cycles is generally large,
$N\gg1$ (see equation [\ref{numcycles}]), and it is unlikely that
stellar mass loss can be exactly described by a model with constant
$\undex$ for a precise number $N$ of cycles (and then stops abruptly).
As result, we are unlikely to know where the final value of $f$ lies
between the turning points. Additional planets, or other perturbations,
increase this uncertainty (see Section \ref{sec:lyapunov}).  In this
case of incomplete information, we can write the mean value of the
energy (averaged over the cycles) in the form 
\be 
2 u^2 \energy = - E + {\gamma^2 x^2 \over E^2} \, ,
\label{enermean} 
\ee
where we have replaced $f$ with the value $1/E$ of its effective
semimajor axis, and where the remaining term averages to zero. This
estimate for the final energy has an uncertainty --- a range of
possible values --- due to the lack of knowledge of where the planet
lies in its orbit during the final cycle. This range of energy is
given by the form 
\be
{\Delta \energy \over \energy} = \pm {2 \gamma x 
(2f - \angmom - E f^2 )^{1/2} \over 2 u^2 \energy} 
\label{enerdelta} 
\ee
$$
= \pm 
{2 \gamma x e_\ast \over E^{3/2} - \gamma^2 x^2 / E^{3/2}} \,,  
$$
where $e_\ast$ is defined by equations (\ref{zelements}) and
(\ref{estar}).  With the energy specified, the final value 
$a_f$ of the semimajor axis is given by 
\be
a_f = - {1 \over 2u \energy_f} \, .
\label{afinal} 
\ee
The expected value of the energy $\energy_f$ is given by equation
(\ref{enermean}), but it can take on any value in the range defined by
equation (\ref{enerdelta}).  Similarly, the final value $e_f$ of the
orbital eccentricity is given by 
\be
e_f^2 = 1 + \angmom (2 u^2 \energy_f) \approx 
1 - \angmom E + \angmom \gamma^2 x^2 / E^2 \, , 
\label{efinal} 
\ee
where the second (approximate) equality holds for the mean value of
energy given by equation (\ref{enermean}). Since the energy $\energy_f$ 
can have a range of values, given by equation (\ref{enerdelta}), the
orbit has a corresponding range of possible eccentricities.

The expressions derived above for the final orbital elements are
expected to be valid provided that the time scale ratio $\tratio$ is
small compared to unity (and hence $|J|\ll1$). The time evolution of
the orbital elements is illustrated in Figure \ref{fig:elements} for a
representative system with mass loss index $\undex$ = 2 and mass loss
parameter $\gamma = 10^{-4}$. Numerical integration of the full
equation of motion (solid curves) show that the semimajor axis and
eccentricity both oscillate and (on average) grow with time. (Note
that the Figure plots $\log[a]$.) The values of the elements $(a,e)$
calculated from the average energy (from equation [\ref{enermean}])
provide a good approximation to the mean evolution of the orbital
elements (see the central dotted curves in Figure \ref{fig:elements}).
Furthermore, using the range of allowed energy calculated from
equation (\ref{enerdelta}), we can calculate upper and lower limits to
the expected behavior of the semimajor axis and eccentricity (shown as
the upper and lower dotted curves). Note that the solutions for $a$
and $e$ oscillate back and forth between these limiting curves. In
this example, the planet becomes unbound near $u = 28$. Prior to that
epoch, near $u \approx 20$, the limits on the energy allow for the
planet to become unbound, and the upper limit for the semimajor axis
approaches infinity. The approximation scheme thus breaks down at this
point. 

\subsection{Numerical Results} 
\label{sec:numerical} 

\begin{figure} 
\centerline{ \epsscale{0.90} \plotone{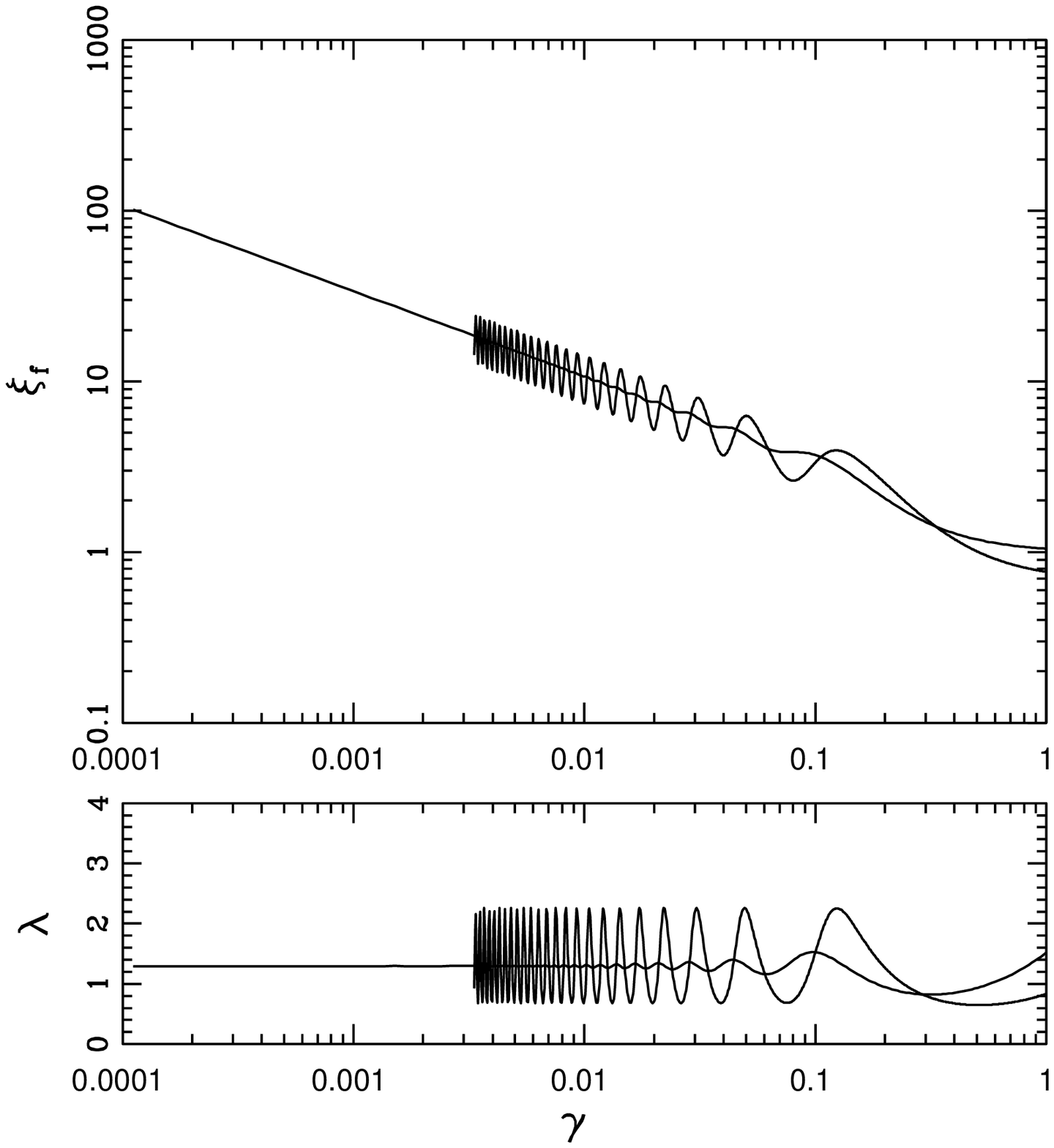} } 
\figcaption{Radial coordinate of the planet at the moment when 
the system becomes unbound, shown here as a function of the mass 
loss parameter $\gamma$ for exponential mass loss (top panel).
The nearly monotonic curve shows the result for the case of 
circular starting orbits; the oscillating curve shows the result
for angular momentum parameter $\angmom$ = 0.9 (which corresponds 
to starting eccentricity $e$ = $\sqrt{0.1} \approx 0.316\dots$). 
Bottom panel shows the value of the parameter $\tratio$ when the planet 
becomes unbound, for both circular starting orbits (smooth curve) 
and eccentric orbits ($\angmom$ = 0.9; oscillating curve). All orbits
are started at periastron. }
\label{fig:endvgam} 
\end{figure}

The equations of motion can be numerically integrated to find the
value $\xi_f$ of the radial coordinate when the system becomes unbound
(when the energy becomes positive). For the case of exponential mass
loss, $\undex$ = 1, the result is shown in Figure \ref{fig:endvgam} as
a function of the mass loss parameter $\gamma$ (top panel). The figure
shows curves for initially circular orbits ($\angmom$ = 1, smooth
curve) and for nonzero starting eccentricity ($\angmom$ = 0.9, rapidly
oscillating curve). Note that the $\angmom$ = 0.9 curve is shown only
for $\gamma > 0.003$; for smaller values of $\gamma$, the curve oscillates
more quickly as a function of $\gamma$, and the curve would appear as
a solid black band in the figure. The bottom panel shows the value of
$\tratio$, the ratio of the orbital period to the mass loss time
scale, evaluated when the system becomes unbound. As expected, the
parameter $\tratio$ is of order unity when the system energy becomes
positive and the planet becomes unbound. For the case of circular
orbits at the initial epoch ($\eta$ = 1), the value of $\tratio \sim$
1.3 for small $\gamma$. For starting orbits with nonzero eccentricity,
the value of $\tratio$ takes on a range of values, but remains of
order unity. For the case shown ($\eta$ = 0.9, oscillating curve), the
parameter $\tratio$ varies between about 1/2 and 2.

\begin{figure}  
\centerline{ \epsscale{0.90} \plotone{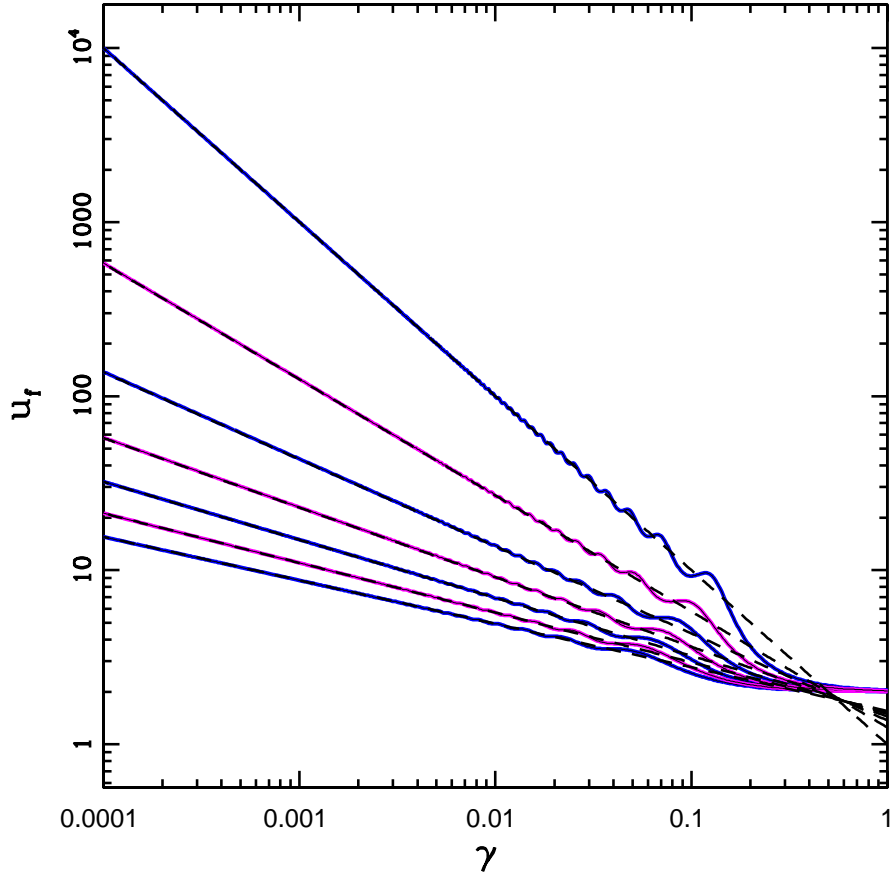} } 
\figcaption{Value of the time/mass variable $u=1/m$ at the moment when 
the system becomes unbound, shown here as a function of the mass loss
parameter $\gamma$, for varying values of the index $\undex$. All of
the curves correspond to circular starting orbits with $\angmom$ = 1 
(which corresponds to starting eccentricity $e$ = 0). The curves
correspond to values of $\undex=0$ (top curve) to $\undex=3$ (bottom
curve). Dashed (black) lines correspond to the fitting function of 
equation (\ref{ufinalfit}). }
\label{fig:ufinalbeta} 
\end{figure}

The above trend holds over a range of values for the mass loss index
$\undex$. Figure \ref{fig:ufinalbeta} shows the value of the time/mass
variable $u=1/m$ when the planet becomes unbound as a function of the
mass loss parameter $\gamma$. Results are shown for mass loss indices
in the range $0\le\undex\le3$. For all values of the index $\undex$,
the curves become nearly straight lines in the log-log plot for small
values of $\gamma$, which indicates nearly power-law behavior of the
form $u_f \sim \gamma^{-p}$, where the index $p=1/(\undex+1)$. We can
find a simple fitting function for the final value $u_f$ of the
inverse mass variable (for initially circular orbits) as a function of
$\gamma$: 
\be
u_f = \left[ 1 + {c_0 \undex \over 1 + \undex} \right] 
\gamma^{-1/(1+\undex)} \, , 
\label{ufinalfit} 
\ee
where $c_0$ is a constant. For $c_0$ = 0.74212, the error is less than
0.4\% for mass loss indices in the range $0 \le \undex \le 4$.
Power-law fits resulting from equation (\ref{ufinalfit}) are shown as
the dashed lines in Figure \ref{fig:ufinalbeta}. The power-law fits
provide a good approximation for $\gamma \lta 0.1$. For larger values
of $\gamma \gta 0.1$, the final value $u_f \to 2$. Equation
(\ref{ufinalfit}) describes the final mass values (for initially
circular orbits) for realistic mass loss prescriptions: Most of the
stellar mass is usually lost on the asymptotic giant branch (AGB), 
where the mass loss function has $\undex \sim$ {\sl constant} (see
Fig. 13 of Veras et al. 2011). Note that mass loss on the AGB takes
place through a series of pulses, but this complication primarily
affects tides (Mustill \& Villaver 2012), rather than the overall 
mass loss profile. 

\begin{figure} 
\centerline{ \epsscale{0.90} \plotone{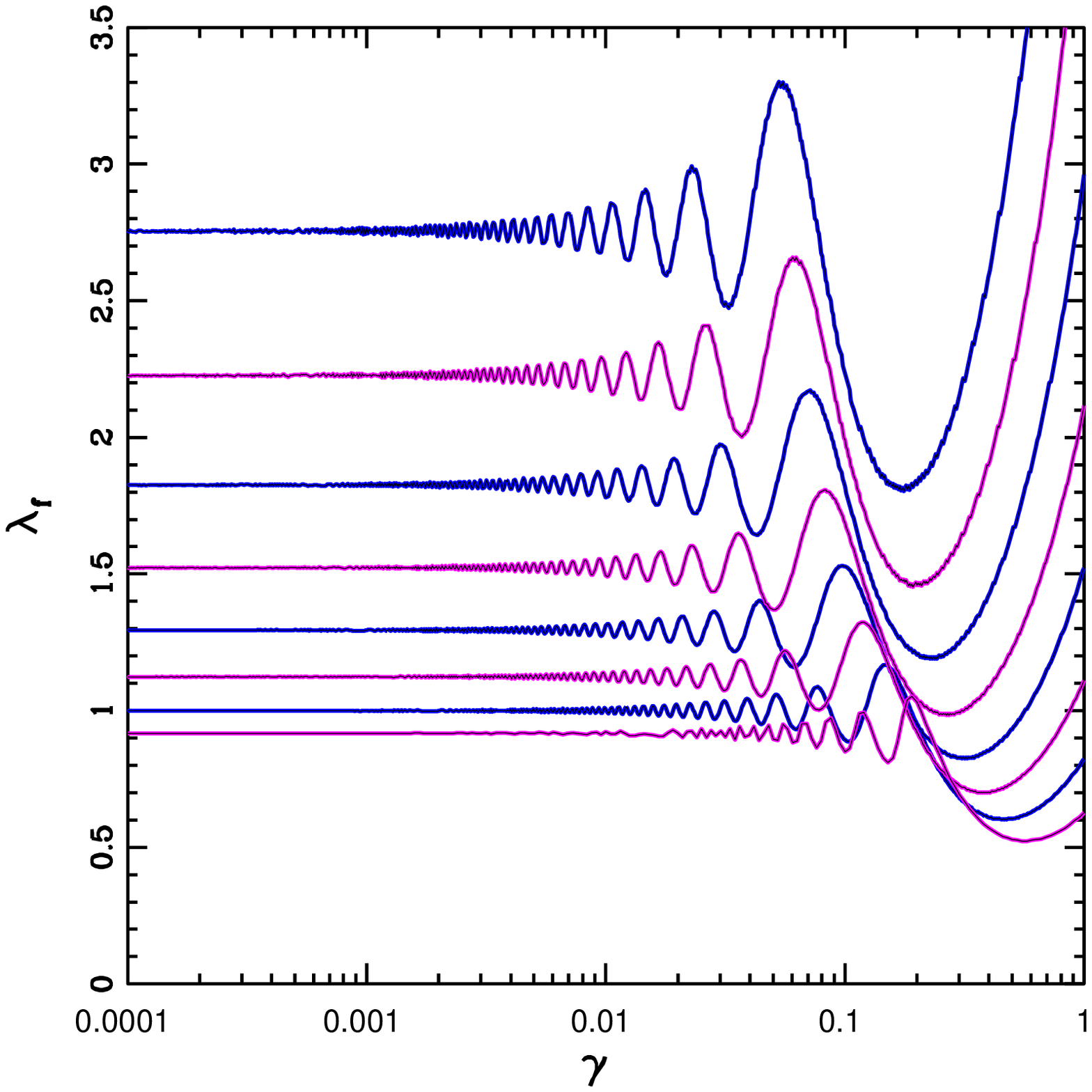} } 
\figcaption{Value of the time scale ratio $\tratio$ evaluated at 
the moment when the system becomes unbound, shown here as a function
of the mass loss parameter $\gamma$, for varying values of the index
$\undex$. All of the curves correspond to circular starting orbits
$\angmom$ = 1 (which corresponds to starting eccentricity $e$ =
0). The curves correspond to values of $\undex=-0.5$ (bottom curve)
to 3.0 (top curve). } 
\label{fig:lambdabeta} 
\end{figure}

A related result in shown in Figure \ref{fig:lambdabeta}, which plots
the values of the ratio $\tratio$ of time scales, evaluated at the
moment when the planet becomes unbound, as a function of the mass loss
parameter $\gamma$. In the limit of small $\gamma$, the time scale
ratio $\tratio$ approaches a constant value (of order unity). The
finding that $\tratio$ has a value of order unity (in the limit of
small $\gamma$) when the planet becomes unbound is expected: In
physical terms, this result means that the mass loss time scale has
become shorter than the orbital period, so that the potential well
provided by the star is changing fast enough that the orbital motion
does not average it out.

\begin{figure}  
\centerline{ \epsscale{0.90} \plotone{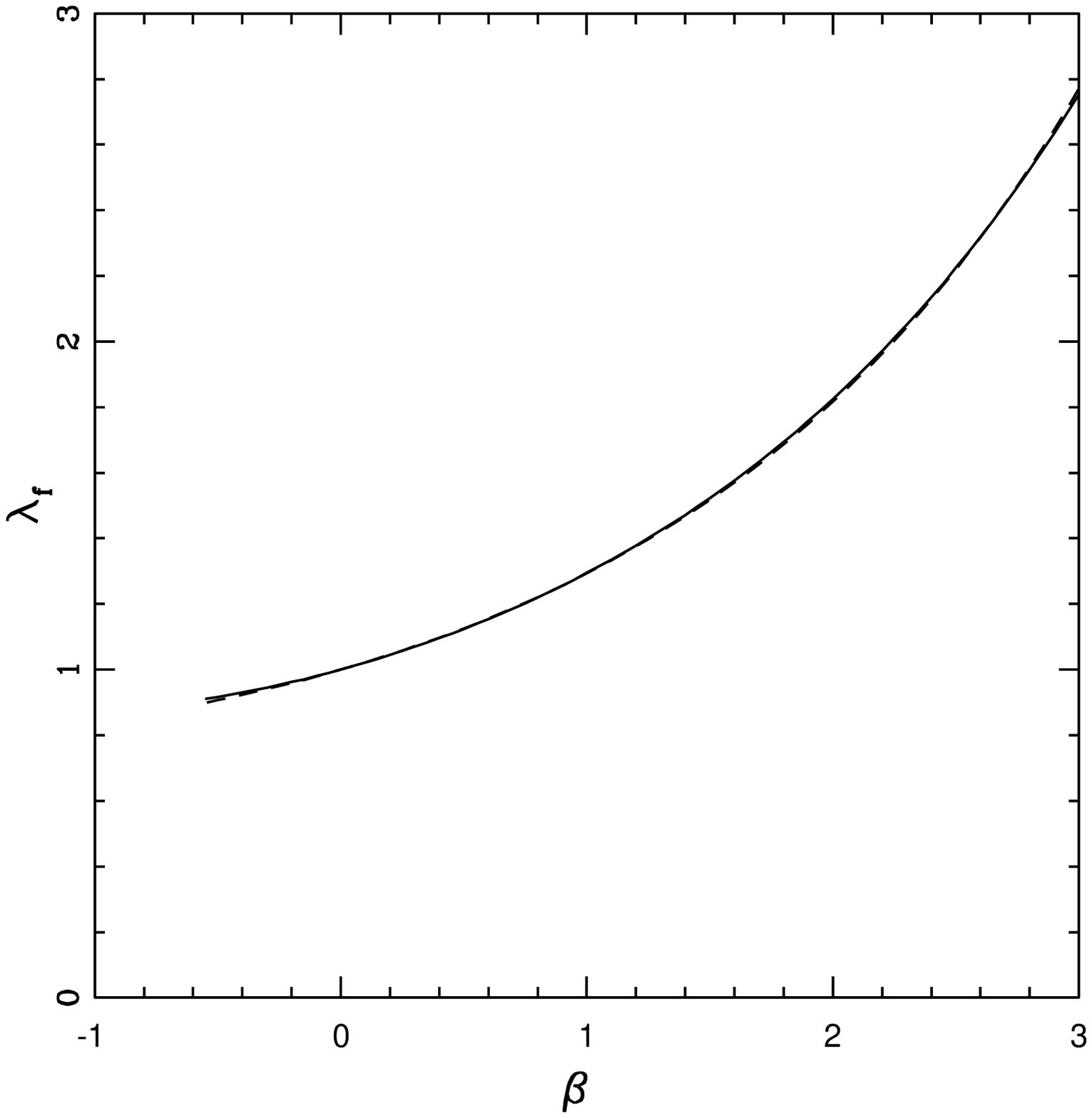} } 
\figcaption{Value of the time scale ratio $\tratio$ evaluated at 
the moment when the system becomes unbound, shown here as a function
of the mass loss index $\undex$ which defines the time dependence of 
stellar mass loss. These values correspond to the limit of small 
mass loss parameter $\gamma \to 0$ and the limiting case where the 
eccentricity of the starting orbit $e=0$. The dashed curve, which is
nearly identical to the solid curve, shows a simple fit to the
function $\tratio_f(\undex)$, as described in the text. As shown in 
the text, $\tratio_f$ = 1 for the particular case $\undex$ = 0. } 
\label{fig:lambdalimit}
\end{figure}

The limiting values of the time scale ratio $\tratio$ are shown in
Figure \ref{fig:lambdalimit} as a function of the mass loss index
$\undex$. Here the time scale ratios are evaluated at the moment when
the planet becomes unbound. Results are shown for the limiting case of
small $\gamma$ (from Figure \ref{fig:lambdabeta} we see that the time
scale ratio $\tratio$ approaches a constant value as $\gamma\to0$).
All of the values are of order unity; for the particular case where
$\undex$ = 0, the final value of the time scale ratio $\tratio$ = 1.
Since this function $\tratio_f(\undex)$ is useful for analysis of orbits
in systems losing mass, we provide a simple fit. If we choose a fitting 
function of the form 
\be
\log \tratio_f = c_1 \undex + c_2 \undex^2 \,, 
\label{fit} 
\ee  
where $c_1$ and $c_2$ are constants, we obtain a good fit with the
values $c_1$ = 0.21658 and $c_2$ = 0.04102.  The fitting function is
shown as the dashed curve in Figure \ref{fig:lambdalimit}, and is
almost indistinguishable from the numerically integrated solid curve.

\section{Lyapunov Exponents for Two Planet Systems with Mass Loss} 
\label{sec:lyapunov}  

The discussion thus far has focused on single planet systems, whereas
many solar systems contain multiple planets. In order to see how
multiple planets affect orbital evolution during mass loss, we
generalize the treatment to study systems consisting of two planets
and a central star with decreasing mass.  Such a 3-body configuration
represents a crude model for our Solar System, where the motions of
only the three most dominant objects (Jupiter, Saturn and the Sun) are
considered. As a starting point, we fix the planetary masses and
initial orbital elements $(a,e)$ to those of Jupiter and Saturn, and
set the initial stellar mass to $M_{0\ast}$ = 1.0 $M_{\odot}$.  We
also restrict the orbits to a plane, thereby reducing the number of
phase space variables from 18 to 12. To start, the mass loss function
is taken to be an exponential model with index $\undex$ = 1, although 
this law is generalized later. 

\begin{figure}  
\centerline{\epsscale{0.90} \plotone{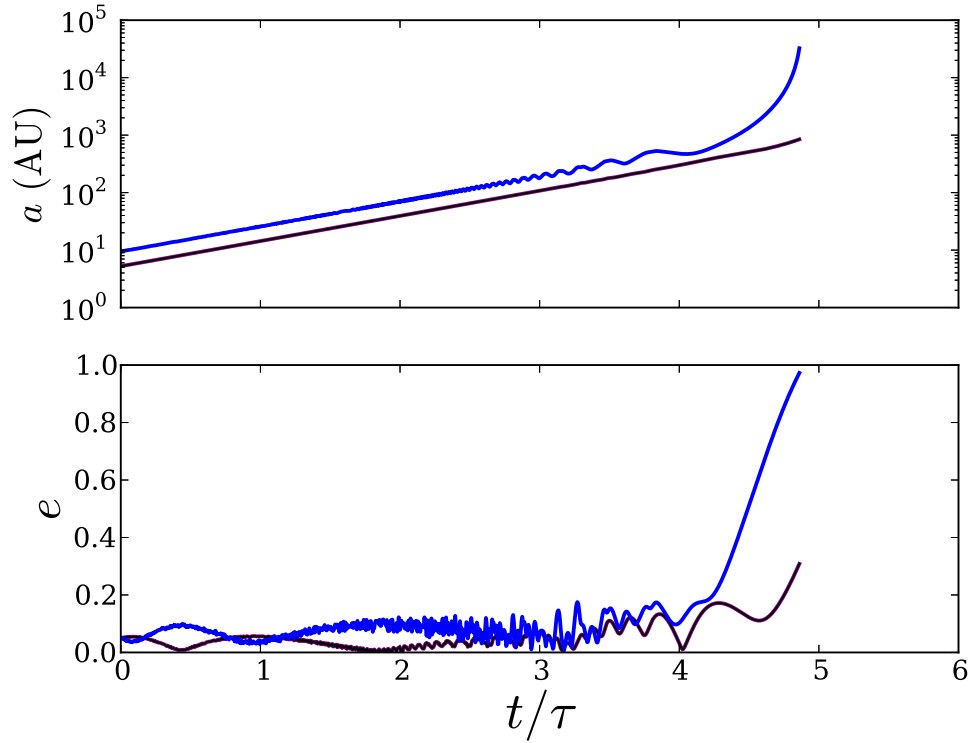} } 
\figcaption{Osculating semimajor axis and orbital eccentricity for a
pair of planets orbiting an initially solar-mass star with mass loss
time scale $\tau = 10^5$ years.  Planets have masses, initial
semimajor axis and eccentricities of Jupiter and Saturn.  The orbital
elements evolve in a roughly predictable manner, with the semimajor
axes increasing smoothly and the eccentricities oscillating on secular
time scales, but remaining relatively constant until the star has
lost the majority of its initial mass.}
\label{fig:orbelements}
\end{figure}

As long as the system suffers no close encounters, the orbit of an
individual planet is similar to that described by the variable mass
two-body problem. An example of the evolution of the osculating
orbital elements $(a,e)$ for our benchmark system (see above) is shown
in Figure \ref{fig:orbelements} for a mass loss time scale of $10^5$
yr.  As each planet orbits in an outward spiral, the semimajor axis
increases approximately exponentially in time (in inverse proportion
to the stellar mass). The eccentricity oscillates rapidly on orbital
time scales, and more slowly on secular time scales (as the planets
exchange angular momentum), but remains close to its starting value
until stellar mass loss has taken place for a few e-folding times.  
The product of the semimajor axis and stellar mass ($aM_\ast$) is 
approximately constant until a few e-folding times have elapsed.
After a critical amount of mass is lost, the orbital elements
$a\to\infty$ and $e\to1$, and planets can become unbound. Notice that
at this point, the stellar mass is only a few percent of the initial
value; as a result, this scenario is rather artificial for stars like
our Sun, which are only expected to lose about half of their initial
masses.  However, larger stars lose a greater fraction of their
original masses. For example, a star with initial mass $M_{0\ast}$ 
$\approx 8 M_{\odot}$ is expected to end its life as a white dwarf
with roughly $\sim 15\%$ of its original mass (where the final mass
fraction depends on the stellar metallicity).

\begin{figure} 
\centerline{\epsscale{0.90} \plotone{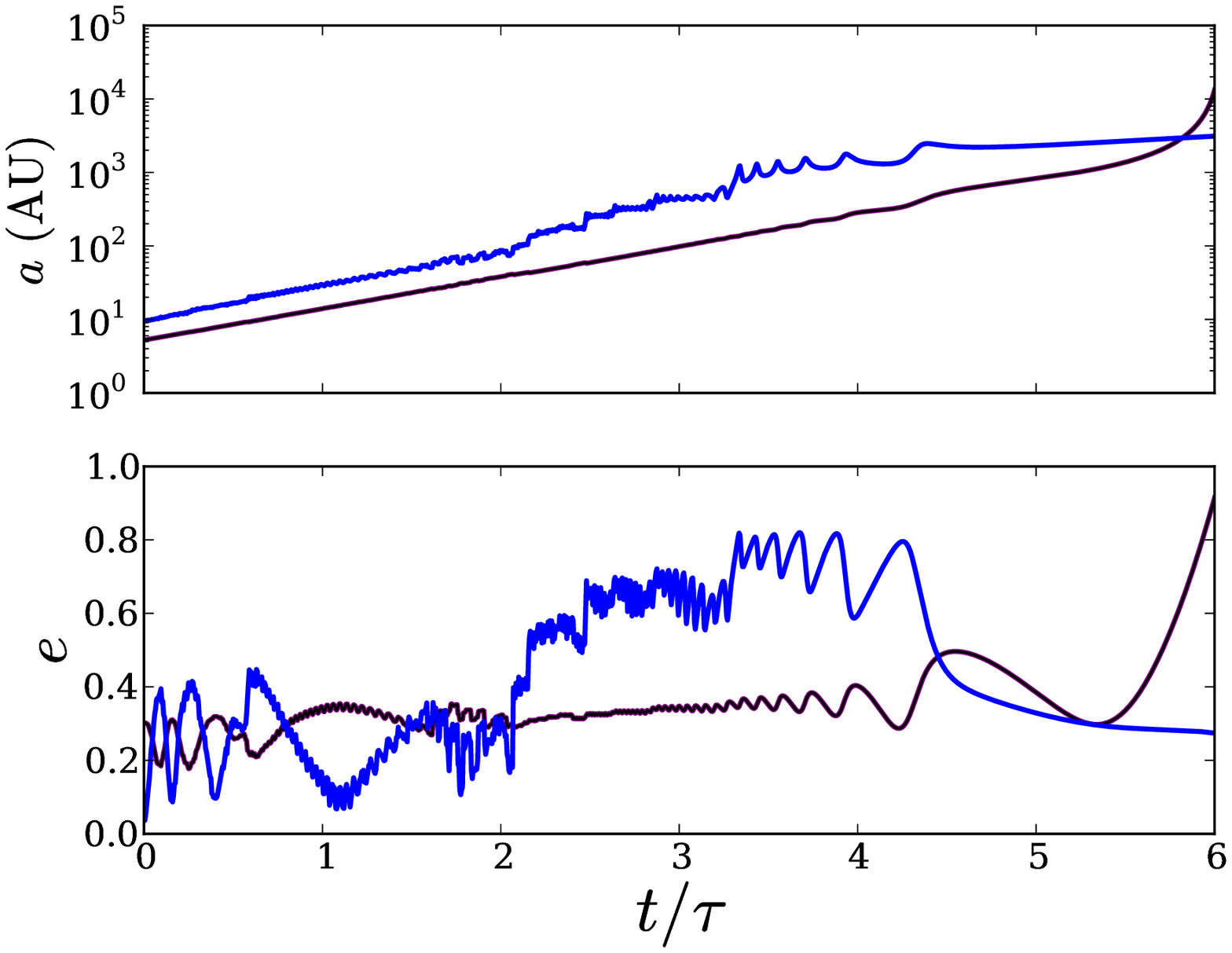} } 
\figcaption{Same as Figure \ref{fig:orbelements}, but with the initial  
eccentricity of Jupiter increased to $e_0 = 0.3$.  This increase in
eccentricity allows for orbital crossings and increases chaotic
behavior.}
\label{fig:orbchaotic}
\end{figure}

The evolution of the orbital elements can differ dramatically if the
planets reach a small enough separation so that orbital crossings can
occur.  In this regime, chaos dominates and the orbital elements
evolve in a less predictable manner.  An example is shown in Figure
\ref{fig:orbchaotic} for the same parameters used in Figure
\ref{fig:orbelements}, but with the initial eccentricity of the inner
planet (Jupiter) increased to $e=0.3$ (a typical value for the current
exoplanet sample).  Although the system initially exhibits nearly
periodic behavior, by the time $t=\tau$ this stable evolution has been
compromised.

Studies of dynamical systems generally use the maximum Lyapunov
exponent as an indication of the level of chaos present
\citep[e.g.,][]{lichtenberg,strogatz}. If a system is chaotic, two
nearby trajectories in phase space initially differing by a small
amount $\delta_0$ should diverge according to 
\be
\delta (t) = \delta_0 e^{\growth t} \qquad {\rm with} \qquad  
\growth > 0. 
\ee
The Lyapunov time is thus $\tlyap=1/\growth$. The long-term dynamical
stability of the solar system has been explored in the absence of
stellar evolution \citep{batygin,laskar}, and current estimates of the
Lyapunov time for the solar system (while the mass of the Sun remains
constant) are $\tau \approx 5$ Myr \citep{sussman}, but this value 
decreases when stellar mass loss is introduced, as demonstrated below.

\begin{figure} 
\centerline{ \epsscale{0.90} \plotone{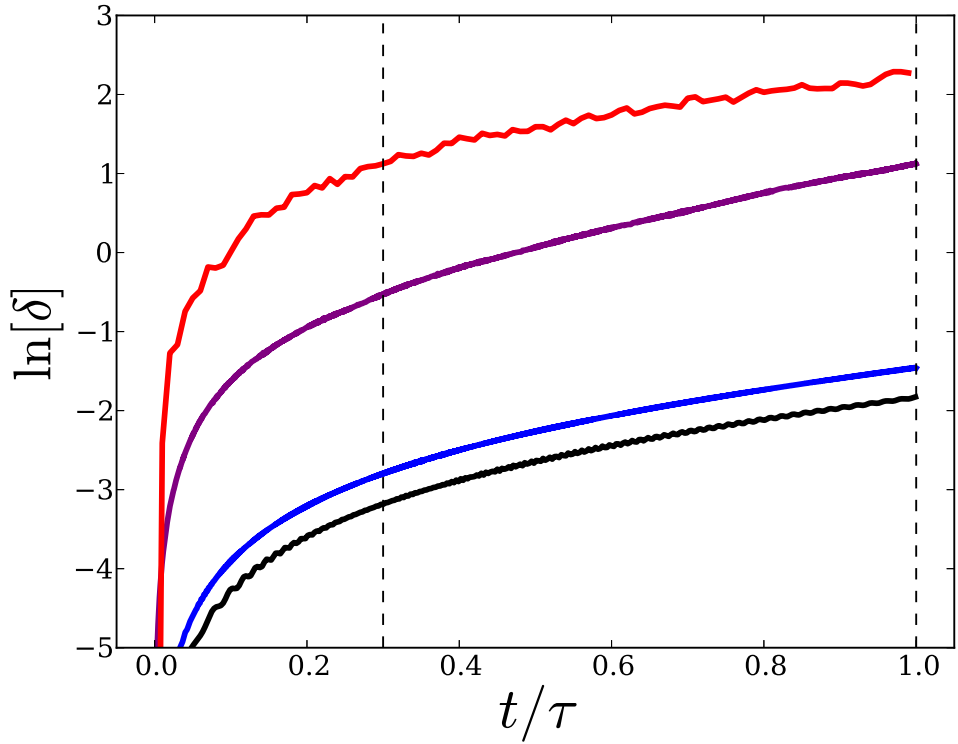} } 
\figcaption{Curves showing the divergence of two trajectories separated 
by a small distance $\delta_0$ for different values of the mass loss 
time scale $\tau$.  Black curve is $\tau = 10^4$; blue is $\tau =
10^5$; purple is $\tau = 10^6$; and red is $\tau = 10^7$ (yr).  After
a period of initial growth, the trajectories diverge exponentially,
indicated by the linear shape of the latter portions of the graphs.
The regions between the dashed lines were used to calculate the
Lyapunov exponent.  Note that the time variable has been scaled by the
mass loss time scale $\tau$.}
\label{fig:delta}
\end{figure}

Here we determine the Lyapunov times as a function of the mass loss
time scale via numerical integrations.  We define a ``real'' system
along with a ``shadow'' system where the initial conditions of the
shadow system differ by a small amount $\delta_0$.  By integrating
both systems simultaneously and monitoring the quantity $\delta(t)$,
we can calculate the divergence of the neighboring trajectories and
then estimate the maximum Lyapunov exponent. Since a three-body system
restricted to a plane consists of 12 phase space variables, there is
some choice in defining the quantity $\delta$. For the sake of
definiteness, we define $\delta$ according to 
\be
\delta = \sqrt{(x_r - x_s)^2 + (y_r - y_s)^2},
\label{delta}
\ee 
where $(x,y)$ are Cartesian coordinates for a planet's location, and
where the subscripts $r$ and $s$ refer to the ``real'' and ``shadow''
trajectories respectively. Since chaotic systems display complicated
behavior, the functions $\delta(t)$ will vary for effectively
equivalent cases. As a result, for each system of interest, we run
1000 cases with both a ``real'' and a ``shadow'' system.  To extract
the Lyapunov exponent, we can either average together the 1000 runs to
construct a single function $\delta(t)$ and use the result to find the
exponent, or, we can find the exponent from each of the 1000
individual cases and then find the average exponent. Both schemes
produce the same values; here we present results for the former case.
Figure \ref{fig:delta} shows an example of the time evolution of
$\delta(t)$ for different values of the mass loss time scale $\tau$.
After an initial period of transient growth (roughly delimited by 
$t \lta 0.3 \tau$), the divergence metric $\delta(t)$ increases
exponentially with time and the Lyapunov exponent can be obtained 
by finding the slope of the line defined by $\ln{\delta}$ = 
$\ln{\delta_0+\growth t}$.

For each value of the mass loss time scale $\tau$, the maximum
Lyapunov exponent was calculated as outlined above.  Since the maximum
possible separation between the reference and shadow systems is finite
(using the definition of $\delta$ in equation [\ref{delta}]), the
curves of growth eventually saturate.  Thus, to extract the Lyapunov
exponent $\growth$, we want to measure the curves of divergence after
the initial interval of transient behavior but before saturation
occurs.  In most cases, $\growth$ was calculated from the time-series
data for times $\tau/3 \leq t \leq \tau$; this time interval is
delimited by the vertical dashed lines in Figure \ref{fig:delta}. An
exception was made for the case of extremely slow mass loss, however,
where $\tau$ = 100 Myr. For this scenario, the time scale for mass
loss is longer than the ``natural'' Lyapunov time of $\sim 10$ Myr
(the value obtained without mass loss), and the curves of divergence
saturate before $t = 0.3 \tau$.  In this case, $\growth$ was
calculated only for time series data with $t < 10$ Myr. Notice that
the curves shown in Figure \ref{fig:delta} are not perfectly straight
in the region between the dashed lines; this curvature introduces some
uncertainty in the specification of the Lyapunov time scales. To
estimate this uncertainty, we have calculated the Lyapunov time values
for a wide range of choices for the time intervals. This procedure
implies an uncertainty of a few percent. 

\begin{figure} 
\centerline{ \epsscale{0.90} \plotone{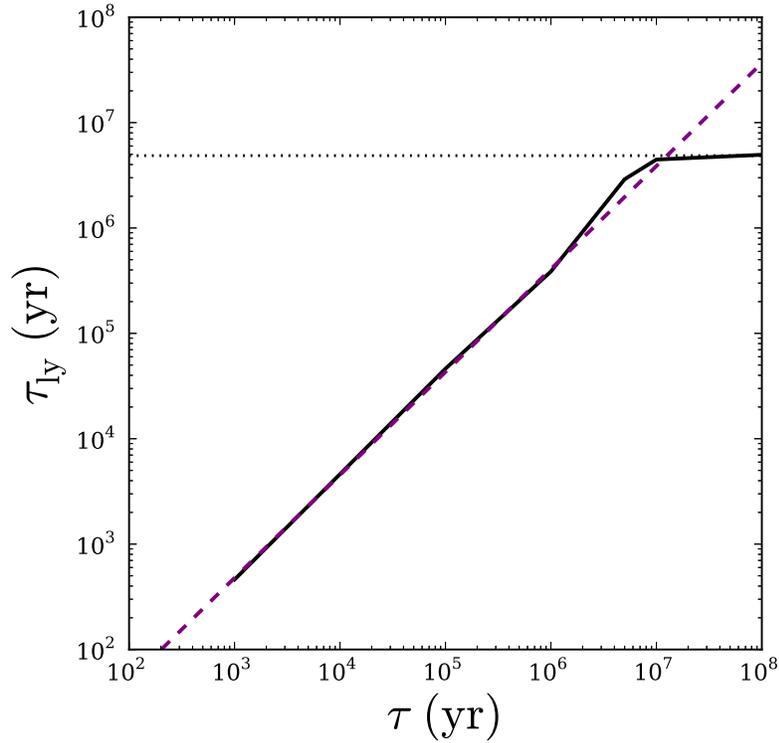} } 
\figcaption{Calculated Lyapunov times $\tlyap$ versus mass loss 
time scales $\tmass$ for a star losing mass exponentially in time 
(mass loss index $\undex$ = 1). The calculated quantities are shown 
(solid curve) along with a least-squares fit for $\tmass\leq10^7$ yr 
(dashed line).  As $\tmass\to\infty$, the Lyapunov time approaches
that of the solar system with constant stellar mass ($\tlyap\sim5$ 
Myr, as marked by the horizontal dotted line).} 
\label{fig:lyapvtau}
\end{figure}

Figure \ref{fig:lyapvtau} shows our numerical values of the Lyapunov
times $\tlyap$ as a function of the mass loss time $\tmass$.  We
performed the analysis described above for each of the two planets in
the system separately and averaged the results.  The horizontal dotted
line -- included here for reference -- corresponds to our numerically
determined Lyapunov time for the model solar system in the absence of
stellar mass loss.  This value is in relatively good agreement with
previous calculations for the complete Solar System \citep{sussman},
but differs slightly because our model considers only two of the four
giant planets. Note that as $\tmass\to\infty$, the Lyapunov time
approaches the dotted line, i.e., the value expected with no mass
loss.  The solid curve shows the calculated values of the Lyapunov
time, whereas the dashed line indicates the least-squares fit (where
the fit was taken over range of mass loss time scales
$\tmass\leq10^7$).  Notice that the slope of this line is close to
unity.  More specifically, we obtained 
\be 
\tlyap \sim \tmass^p 
\qquad {\rm where} \qquad p = 0.99.  
\ee 
Note that this fitted line cannot be meaningfully extrapolated below
$\tmass=10^2-10^3$. In this regime, the mass loss time scale $\tmass$
becomes comparable to the orbital periods of the planets, and the
dynamics of even single-planet systems becomes complicated (see the
previous section).

As a consistency check, we also explored other choices for the metric
$\delta$ that measures the difference between nearby trajectories.  An
especially compelling option is to use the semimajor axis $a$, because
unlike the physical distance between the reference and shadow
trajectories, there is no upper limit on this quantity. The previous
calculation was thus repeated using
\be
\delta = |a_r - a_s|.
\ee
Our results are similar, however, which indicates that the Lyapunov
times do not depend sensitively on the choice of $\delta$.

\begin{figure} 
\centerline{ \epsscale{0.90} \plotone{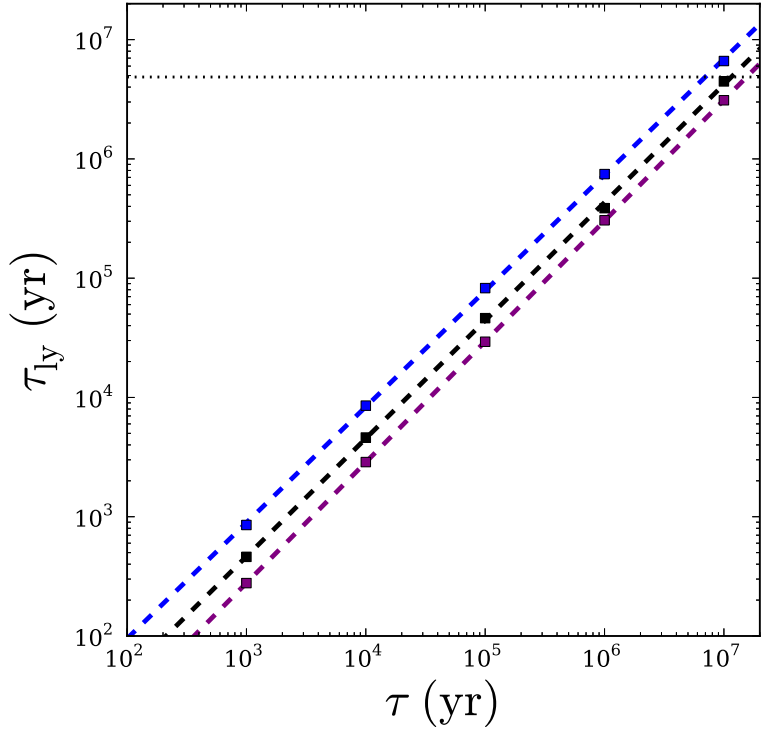} } 
\figcaption{Lyapunov time scale $\tlyap$ versus mass loss time scale
$\tmass$ for systems where the star loses mass through three different 
decay laws. From top to bottom, the blue curve shows the results for
$\undex=0$; the black curve shows the results for exponential mass
loss ($\undex=1$); and the purple curve shows the results for constant
mass loss rate ($\undex=2$). For all three cases, the square symbols
show the numerically calculated time scales and the dashed lines show
a least-squares fit. These three examples show similar behavior, which
indicates that the linear relationship between Lyapunov time scales
and mass loss time scales is largely independent of the particular
mass loss formula. }
\label{fig:lyapvtau_all}
\end{figure}

Next we would like to ensure that the (nearly) linear relation between
the Lyapunov time and mass loss time is not an artifact of the
exponential ($\undex$ = 1) mass loss law that was chosen. Toward this
end, we have explored two additional functional forms for the mass
loss law: The first used vanishing mass loss index $\undex$ = 0 (see
equation [\ref{mtzero}]), whereas the second used a constant mass loss
rate with $\undex=2$ (see equation [\ref{mtconstant}]). For both of
these mass loss functions, we obtained nearly the same power-law
relation for the Lyapnunov time scale versus the mass loss time scale,
i.e., $\tlyap \sim \tmass^p$, where $p = 0.98$ and $p = 1.01$ using
equations (\ref{mtzero}) and (\ref{mtconstant}), respectively. The
results, shown in Figure \ref{fig:lyapvtau_all}, are thus nearly
identical, independent of the index $\undex$ of the mass loss
function. This finding suggests that this power-law trend is robust.

Since the Lyapunov time scale is found to be comparable to the time
scale for mass loss, we generally expect such solar systems to be only
moderately influenced by chaos. In order for chaos to fully erase
initial conditions for a dynamical system, nearby trajectories in
phase space must diverge for several Lyapunov times. For example, in
order for a starting uncertainty of $1^\circ$ in position angle to
grow into $360^\circ$, one needs $\sim6$ Lyapunov times, or, about 6
mass loss time scales.  For exponential mass loss, e.g., this time
interval would result in the star losing 99.75\% of its initial
mass. Since most stars do not lose such a large percentage of their
mass, the effects of chaos are not expected to completely erase the
initial conditions of these systems. Nonetheless, chaos will partially
erase the initial conditions; this trend will affect our ability to
predict the phase of the orbit at the end of mass loss and will thus 
introduce uncertainty into predictions of the final (post-mass-loss) 
orbital elements (see Section 3.4). 

\section{Applications} 
\label{sec:apply} 

To illustrate the efficacy of the results found in the previous
sections, we consider two representative astronomical problems. For
solar type stars, we find an effective outer edge of the solar system,
i.e., the boundary between planetary bodies that remain bound after
the epoch of mass loss and those that escape (Section \ref{sec:edge}).
Next we consider planets that remain bound to white dwarfs, and find
their final orbital elements (Section \ref{sec:wdplanets}).

\subsection{Outer Boundary of the Solar System} 
\label{sec:edge} 

Suppose that a solar-type star, with initial mass $M_{0\ast}$ = 1
$M_\odot$, loses some portion of its mass over a time interval 
$\Delta t$ = 1 Myr, so that the fraction $m_f$ remains afterward.
Since solar type stars lose most of their mass while they are on
either the Red Giant Branch or Asymptotic Giant Branch (see Hurley et
al. 2000), we expect $\undex$ to be in the range 1--3.  For a fixed
time interval $\Delta t$ and arbitrary index $\undex>1$, the mass loss
parameter $\gamma$ is given by 
\be
\gamma = {1 \over \undex - 1} \left[ 1 - m_f^{\undex-1} \right] 
{1 \over \Delta t} \left( {a_0^3 \over G M_{0\ast} } \right)^{1/2} 
\label{eggamma} 
\ee
$$
= 1.6 \times 10^{-7} 
{1 \over \undex - 1} \left[ 1 - m_f^{\undex-1} \right] 
\left( {a_0 \over 1 {\rm AU} } \right)^{3/2} \,, 
$$ 
where $a_0$ is the initial semimajor axis of the orbit. For initially
circular orbits, we can use the results of Section 3.5 to find the
conditions for which planets become unbound. Using equation
(\ref{ufinalfit}) to define the critical value of $\gamma$, and
equating the resulting value to the expression from equation
(\ref{eggamma}), we can solve for the critical semimajor axis $a_c$,
such that orbits with larger values of $a$ become unbound during the
mass loss epoch.  First we define the constant 
\be
A_\undex \equiv (\undex-1)^{2/3}  
\left[ 1 + {c_0 \undex \over 1 + \undex} \right]^{2(1 + \undex)/3} \,.
\ee
The critical semimajor axis $a_c$ is then given by 
\be
a_c = (\Delta t)^{2/3} (G M_{0\ast})^{1/3} A_\undex 
\left\{ {m_f^{1 + \undex} \over 
\left[1 - m_f^{\undex-1} \right]} \right\}^{2/3} 
\ee
$$ 
\approx \, (34,000 {\rm AU}) \, A_\undex 
\left\{ {m_f^{1 + \undex} \over 
\left[1 - m_f^{\undex-1} \right]} \right\}^{2/3} \,.
$$ 
We can thus find the critical semimajor axis $a_c$ for any given value
of the index $\undex$ and the remaining mass fraction $m_f$. Note that 
the resulting value of $a_c$ is a sensitive function of the index
$\undex$. To leading order, $a_c \sim$ (34,000 AU) 
$m_f^{2(1+\undex)/3}$. If we use $m_f$ = 1/2, the value expected for
the Sun, then the critical value of the semimajor axis $a_c \approx$
8500, 5350, and 3370 AU for indices $\undex$ = 2, 3, and 4 (these
values for $a_c$ are in general agreement with the results of Veras \&
Wyatt 2012). For planets that escape, the corresponding velocities are
small, in the range 0.3 -- 0.5 km/s.  For initially eccentric orbits,
planets will become unbound for a wide range of starting semimajor
axes, depending on the initial phase of the orbit (see Section
\ref{sec:results}); however, this range is centered on the mean 
values found here.  

Note that in the limit $\undex \to \infty$, we formally obtain
$a_c\to0$ for any value of the remaining mass fraction $m_f<1$.
However, this formulation of the problem breaks down before that limit
is reached: For large values of $\undex$, even though the time
interval is fixed, the mass loss rate accelerates rapidly so that
most of the mass is lost near the end of the time interval.  Stellar
mass is thus lost (effectively) through a step function in the limit
of large $\undex$. In this limit, the results of Section 3.3 apply, 
and circular orbits become unbound (remain bound) for mass fraction 
$m_f < 1/2$ ($m_f > 1/2$). 

\subsection{Orbital Elements for White Dwarf Planets} 
\label{sec:wdplanets} 

Now consider a progenitor star with initial mass $M_{0\ast}$ = 5
$M_\odot$, which evolves into a white dwarf with mass $M_{wd}$ = 1
$M_\odot$ (Hurley et al. 2000); the final mass fraction $m_f$ = 1/5
($u_f$ = 5). For purposes of illustration, we assume that the mass is
lost over a single epoch that can be described by a single value of
the mass loss index $\undex$, with a time scale $\Delta t$ = 1 Myr.
For orbits with starting semimajor axes $a$ and eccentricities $e$, we
would like to know the final orbital elements, after the epoch of mass
loss. For orbits with starting semimajor axis $a$ in the range $1-100$
AU (closer planets are often accreted by the star), the value of
$\gamma$ falls in the range $\gamma=10^{-7}-10^{-4}$.  The mass loss
parameter $\gamma$ is thus small and nearly independent of the index
$\undex$ (see equation [\ref{eggamma}]). The time scale ratio
$\tratio=\gamma u^{\undex+1} f^{3/2}$. For the systems of interest,
the largest $\gamma$ value is thus $\sim10^{-4}$, the largest value of
$u$ = 5, and the largest value of $\undex$ = 3; the largest value of
the time scale ratio is thus $\tratio \sim 0.063$, so that $\tratio^2
\le 0.004$. In the approximation scheme developed in Section 3.5, we
have exact results when the integral $J$ is small, where $J={\cal
  O}(\tratio^2) \le$ 0.004 (see also Appendix \ref{sec:jbound}).

The initial conditions for a planetary orbit include not only the 
semimajor axis and eccentricity $(a,e)$, but also the phase of the
orbit at $t$ = 0.\footnote{A full specification would also include
  three additional angles, e.g., the longitude of periastron, the
  longitude of the ascending node, and the inclination angle, but we
  can orient the coordinate system to eliminate this complication.}
This latter quantity is specified by the initial value of the
radial coordinate $\xi_0=f_0$, which lies in the range 
$1-e\le\xi_0\le1+e$. With $\xi_0(f_0)$ determined, the integration
constant $E$ is given by equation (63). 

In the limit $J\to0$, the equation of motion shows that the function
$f$ oscillates back and forth between its turning points (given by
equation [\ref{turningpts}]) while the system loses mass (analogous to
the evolution depicted in Figure \ref{fig:limits}).  If the duration
of the mass loss epoch is specified exactly (equivalently, if
$m_f=1/u_f$ is known exactly), then we can determine the final value
of the function $f_f=f(x_f)$, where $x_f=u_f^\alpha$.  Let $N_c$
denote the number of complete cycles that that function $f(u)$
executes during the mass loss phase, where a cycle is defined as 
motion from one turning point to the other (a half orbit). In
addition, the orbit must turn through a partial cycle from its 
starting value $f_0$ to the first turning point $f_j$ (where $j$ =
1,2) and must turn through another partial cycle from the final
turning point $f_k$ (where $k$ = 1,2) to the final value $f_f$. After
integrating the equation of motion (\ref{fxfive}), we thus obtain 
\be
{E^{3/2} \over \pi \alpha \gamma} \left[ 1 - m_f^\alpha \right] 
= (\delta N)_0 + N_c + (\delta N)_f \,, 
\label{ffinal} 
\ee
where the integration constant $E$ is given by equation 
(\ref{edefine}) and we have defined 
\be 
(\delta N)_0 \equiv \pm {E \over \pi} 
\int_{f_0}^{f_j} {f df \over (f-f_1)^{1/2} (f_2-f)^{1/2}} \,, 
\label{nstart} 
\ee
and 
\be 
(\delta N)_f \equiv \pm {E \over \pi} 
\int_{f_k}^{f_f} {f df \over (f-f_1)^{1/2} (f_2-f)^{1/2}} \,, 
\label{nfinal} 
\ee
where the $\pm$ signs are chosen to keep the integrals positive. Note
that equations (\ref{ffinal} -- \ref{nfinal}) completely specify the
final value of the function $f_f$. After solving these equations for
$f_f$, we can use equation (\ref{jzenergy}) to find the final value
$\energy_f$ of the orbital energy, and then use equation
(\ref{afinal}) and (\ref{efinal}) to find the semimajor axis and
eccentricity of the orbit after mass loss has ended.

Although the procedure described above is exact in principle (subject
to the approximation $J\to0$), there exists a problem: In equation
(\ref{ffinal}), the quantities $(\delta N)_0$ and $(\delta N)_f$ are
less than unity by definition, whereas $N_c$ (and the left-hand side
of the equation) is much larger, i.e., $N_c \sim 10^3 - 10^6$ for the
orbits considered here. To find the value $f_f$ necessary to specify
the phase of the final orbit, this approximate description for mass
loss must be correct to better than 1 part in $10^3$ ($10^6$) for
orbits with starting $a$ = 100 AU (1 AU). It is unlikely that the mass
loss function for a real astronomical system obeys this simple model
to such a high degree of fidelity. As result, even though we have an 
exact solution for the model equation, we cannot predict with 
certainty the final phase of the orbit for realistic systems. 

Given the uncertainty outlined above, post-mass-loss orbits can be
described in terms of the expected values of the energy $\energy_f$,
semimajor axis $a_f$, and eccentricity $e_f$, as well as the possible
ranges of values for the orbital elements (given the range of phases).
For the sake of definiteness, we assume that the orbit starts at
periastron (at the start of the mass loss epoch) and that the mass
loss index $\undex$ = 3. The orbit has starting angular momentum
$\angmom = 1 - e^2$. The integration constant $E$ is then given by
$E=1-\gamma^2(1-e)^2$ (from equation [\ref{edefzero}]) and the energy
$\energy_f$ at the end of the mass loss epoch (from equation
[\ref{enermean}]) becomes 
\be
\energy_f = {1 \over 50} 
\left\{-1 + 390,625 \gamma^2 + (1-e)^2 \gamma^2 + 
{\cal O} (\gamma^4) \right\} \, . 
\ee
Note that we can ignore the second $\gamma^2$ term in the above
expression. The expected final value of the (dimensionless) semimajor
axis (from equation [\ref{afinal}]) is given by 
\be
a_f = 5 \left[1 - 390,625 \gamma^2 \right]^{-1} \,,
\label{afinalmean} 
\ee
and the corresponding expected value of the eccentricity (from 
equation [\ref{efinal}]) is given by 
\be
e_f^2 = e^2 + 390,625 \angmom \gamma^2 \,,
\ee
where we work to the same order of approximation as for $a_f$ 
(recall that $e$ is the starting, pre-mass-loss value of the
eccentricity). Since $\gamma\sim10^{-7}(a_0/1\rm{AU})^{3/2}$, 
the correction term is small: 390,625 $\gamma^2 \sim 4 \times$
$10^{-9}(a_0/1\rm{AU})^{3} \le 0.004$ since $a_0 \le 100$ AU.
The mean value of final semimajor axis is thus about 5 times the
starting value, as expected, and the leading order correction has been
quantified. The square of the eccentricity increases by a similar
amount. Because of the range of possible orbital phases at the end 
of mass loss, the orbital elements can differ from these mean values 
according to the relations 
\be
{\Delta a_f \over a_f} = {\Delta (e_f^2) \over e_f^2} = 
{\Delta \energy_f \over \energy_f} \approx \pm 
{1250 \gamma e_\ast \over E^{3/2}} \,,
\ee
where we have used equation (\ref{enerdelta}) and where
$e_\ast^2=e^2+\gamma^2(1-e)^2$ is the effective eccentricity of the
function $f(u)$ during the epoch of mass loss. Thus, the total
(relative) width of the distribution of possible final orbital
elements is thus $\sim2500\gamma{e}$. Note that this range is often
larger than the correction to the mean values. Consider a planet with
starting semimajor axis $a_0$ = 100 AU and eccentricity $e$ = 0.30.
The mean value of the final semimajor axis is $a_f \approx 502$ AU,
only 2 AU larger than the value suggested by the simple scaling law
$aM_\ast\approx$ {\sl constant} that is often used. However, the range
of possible values about this mean is about $\Delta{a_f}=\pm19$ AU. 
Similarly, the final eccentricity has mean value $e_f \approx 0.306$,
and the width of the range is about $\pm 0.006$.

\section{Conclusion} 
\label{sec:conclude} 

\subsection{Summary of Results} 

This paper has reexamined the classic problem of the evolution of
planetary orbits in the presence of stellar mass loss. Although this
issue has been addressed in previous studies (see Section
\ref{sec:intro}), we generalize existing work to include a new
analytic formulation for time-dependent mass loss and to determine
Lyapunov time scales for multiple planet systems. In particular, we
consider a class of model equations where the mass loss index $\undex$
is constant (see equation [\ref{defindex}]), which allows for a wide
range of time dependence for the mass loss rates and allows for a
number of new results to be obtained analytically. Our main results
can be summarized as follows:

[1] Previous numerical studies show that planetary orbits often obey
the approximate law $a{m}\approx$ {\sl constant}, where $a$ is the
semimajor axis, and where this approximation holds as long as the time
scale for mass loss is significantly longer than the orbital period.
By writing the equation of motion in the form given by (\ref{fuconst})
and (\ref{tratiocon}), we show analytically why this law holds (see
equations [\ref{enerfull} -- \ref{amconstant}]). In addition, the
differential equation (\ref{energydq}) for the energy $\energy$ shows
that the energy is a strictly increasing function of time, so that the
semimajor axis (defined via $a \sim 1/|\energy|$) increases
monotonically (whereas the orbital radius $\xi$ oscillates in and
out).

[2] Previous literature often claims that the orbital eccentricity
remains constant during the early phases of stellar mass loss. In
contrast, this work shows that the eccentricity oscillates back and
forth between well defined limits during the phase of mass loss, and
that the amplitude of these oscillations grow with time (one example
is shown in Figure \ref{fig:elements}). Moreover, the upper and lower
limits of the eccentricity range can be calculated analytically using
equations (\ref{enermean} -- \ref{efinal}). Note that these
oscillations in the eccentricity, while technically correct, result
from assigning orbital elements (which describe ellipses) to orbital
paths that are not elliptical. The actual orbit expands with time,
and, for example, the outer turning point of the orbit increases
monotonically (it does not oscillate).

[3] In the limit of rapid mass loss, $\tratio \to \infty$, we obtain
analytic solutions that describe orbits for the entire class of mass
loss functions (see Section \ref{sec:rapidloss}). The condition for
the planet becoming unbound is given by equation (\ref{masslimit}).
For planets that remain bound, the new orbital elements are given by
equation (\ref{newelements}).

[4] Not all mass loss functions lead to planets becoming unbound
(except, of course, in the extreme case where the stellar mass
vanishes $m\to0$).  The critical value of the mass loss index is
$\undex=-1$, where systems with mass loss characterized by $\undex<-1$
only lose planets in the $m\to0$ limit. Note that systems with the
transition value of the mass loss index $\undex=-1$ were first
considered by Jeans (1924).

[5] For the particular, intermediate value of the mass loss index
$\undex=0$, we can find analytic expressions for the function $f(u)$
and for the final values of the time scale ratio $\tratio_f$ when the
planet becomes unbound (see Section \ref{sec:bzero}). For arbitrary
values of the mass loss index $\undex \ne 0$, this approach can be
generalized to find analytic expressions for $f(u)$ and the orbital
elements of the planet (see Section \ref{sec:generalind}). 
The resulting expressions are approximate, correct to order
${\cal O} (\tratio^2)$, and are thus accurate over most of the 
mass loss epoch.

[6] One way to characterize the dynamics of these systems is through
the parameter $\tratio_f$. We define $\tratio$ to be the ratio of
dimensionless mass loss rate to the orbital frequency, and $\tratio_f$
is the value at the moment when the planet becomes unbound. For
initially circular orbits, the parameter $\tratio_f$ is always of
order unity, and approaches a constant value in the limit of small
mass loss rates $\gamma$; further, the value of this constant varies
slowly with varying $\undex$ (see Figure \ref{fig:lambdabeta}).  For
orbits starting with nonzero eccentricity, however, the parameter
$\tratio_f$ can depart substantially from unity and varies
significantly with $\undex$ (compare Figures \ref{fig:bzeroplane} and
\ref{fig:btwoplane}).

[7] For multiple planet systems, we find that the Lyapunov times
decrease in the presence of stellar mass loss, so that chaos should
play a larger role in planetary dynamics as stars leave the main
sequence. In fact, the Lyapunov time scale is proportional to (but
somewhat shorter than) the mass loss time scale over a range of
conditions (see Figures \ref{fig:lyapvtau} and
\ref{fig:lyapvtau_all}).  For a typical mass loss time scale of
$\tmass\sim10^6$ yr, the Lyapunov time is $\sim2-3\times10^5$ yr.
Three different forms for the stellar mass loss function have been
considered, all yielding similar results, which suggests that this 
trend is robust.

\subsection{Discussion} 

In spite of its apparent simplicity, the classic problem of planetary
orbits with stellar mass loss is dynamically rich. Here we discuss two
issues that have been highlighted by this present work:

The use of osculating orbital elements is a standard way to describe
planetary motion. In this scheme, the planet is assigned the Keplerian
orbital elements that it would have if it were moving within a purely
Keplerian potential. In systems with stellar mass loss, however, this
approach might be more misleading than illuminating (see also
Radzievskii \& Gel'Fgat 1957, Hadjidemetriou 1963). Consider, for
example, the osculating eccentricity of the orbit. As the star loses
mass, the eccentricity oscillates (Figure \ref{fig:limits}) with an
amplitude that grows with time. This oscillating eccentricity would
seem to imply motion along an elliptical path, where the shape of the
ellipse cycles back and forth between being more elongated and more
round.  However, this ``oscillation'' of the eccentricity is an
artifact of assigning a Keplerian orbital element ($e$) to orbital
motion that is not Keplerian. During the mass loss epoch, the orbit
has inner and outer turning points, analogous to those of an
ellipse. But the turning points of the actual orbit do not oscillate
-- the orbit smoothly spirals outward (e.g., see Figure 2 of Veras et
al. 2011).

An alternate description of the dynamics can be constructed: Before
the onset of mass loss, the orbit {\it is} a Keplerian ellipse and can
be described by the usual orbital elements $(a,e)$, which provide the
initial conditions for the next stage (along with the phase of the
orbit at $t=0$).  During the mass loss epoch, the orbit is not an
ellipse, but the scaled radial function $f=\xi{m}=\xi/u$ follows a
nearly Keplerian trajectory.  For the particular mass loss function
with index $\undex=0$, this analogy is exact. In this case, the
function $f$ obeys the same equation of motion (\ref{firstint}) as the
radial coordinate $\xi$ in the Kepler problem (albeit with a
nonconventional time variable where $dt\sim{du}/u^2$). The scaled
function $f$ has turning points (equation [\ref{turningpts}]), an
effective semimajor axis $a_\ast$ = $1/E$ (see equation
[\ref{zelements}]), and an effective eccentricity $e_\ast$ (equation
[\ref{estar}]). In the general case where $\undex\ne0$, the scaled
function $f$ executes nearly Keplerian motion: Here the equation of
motion for $f$ has a Keplerian form up to a correction of order 
${\cal O}(J)$ = ${\cal O}(\tratio^2)$, where $\tratio\ll1$ for most of
the evolution (until shortly before the planet becomes unbound). Note
that the effective orbital elements $(a_\ast,e_\ast)$ characterizing
the function $f$ are constant, to ${\cal O}(\tratio^2)$, while the
star loses mass. After the epoch of mass loss, the planet (if it
remains bound) once again enters into a Keplerian orbit, now with
elements $(a_f,e_f)$. The problem can thus be described in terms of
three sets of orbital elements: before $(a,e)$, during
$(a_\ast,e_\ast)$, and after $(a_f,e_f)$ mass loss. But the orbital
elements during mass loss $(a_\ast,e_\ast)$ correspond to orbits of
the scaled function $f$ (the orbit in physical space is not Keplerian). 

Another interesting complication arises: These orbits display a type
of sensitive dependence on initial conditions even in the absence of
chaos. For planets that remain bound after the mass loss epoch, the
final orbital elements $(a_f,e_f)$ depend on the phase of the orbit,
at both the start and the end of the mass loss epoch. However, the
number of orbital cycles during mass loss is large,
$N_c\sim1/\gamma\gg1$, and a precision of one part in $N_c$ is
necessary to specify the final orbital phase (given the initial
phase). The orbit can be calculated to sufficient accuracy to specify
the final orbital elements provided that the starting state, the
duration of stellar mass loss, and the form of the mass loss function
are all known well enough. In practice, however, real astronomical
systems will not follow these particular forms to such high accuracy,
so that the final orbital elements cannot be predicted with certainty.
Instead, the expectation value of the orbital elements can be
predicted, along with the range of possible variation about their mean
values (see also Section \ref{sec:wdplanets}). 

\subsection{Future Work} 

There are many opportunities for future work. Analytic studies can be
taken into two directions. First, the formulation developed here can
be applied to a wide range of astronomical systems, including
predictions of orbital elements for planets that remain bound after
stellar mass loss and predictions of the conditions required for
bodies to become unbound. On the other hand, the analytic treatment
can be developed further to include more general mass loss functions,
multiple phases of mass loss, and higher order approximations for the
conditions under which planets become unbound.

For the calculations of the Lyapunov time scales, this paper has
focused on analogs of our own solar system, and has considered only
the motion of Jupiter, Saturn, and the Sun, since these are the most
gravitationally dominant bodies.  However, future calculations should
also include additional planets and a wider range of starting orbital
elements. In addition, this paper has considered relatively short
integrations, spanning at most only $10-100$ Myr.  Although stellar
mass loss is not expected to continue for longer than 100 Myr, the
increased semimajor axes of the planets will change the overall
dynamics and the decreased stellar mass (relative to the planets)
could lead to an increase in dynamical instabilities. As a result,
longer integrations should be performed, using the lower (constant)
stellar mass and the increased semimajor axes of the planets as input.
The calculations of this paper are limited to coplanar planeteary
systems; future work should explore the effects of different
inclination angles.  Finally, given the diversity exhibited in the
observed sample of extrasolar planets, different planetary
configurations should also be considered. These types of calculations
will help us understand the long term fate of planetary systems in
general and will help direct future observations.

\section*{Acknowledgments} 

We would like to thank Jake Ketchum, Kaitlin Kratter, Dimitri Veras,
and Eva Villaver for useful discussions.  This work was supported by
NSF grant DMS-0806756 from the Division of Applied Mathematics, NASA
grant NNX11AK87G (FCA), and NSF grants DMS-0907949 and DMS-1207693
(AMB).

\bigskip 

\appendix 
\section{Bounds on the $J$ Integral} 
\label{sec:jbound} 

In the limit $|J|\ll1$, we have a completely analytic description of
the dynamics. It is thus useful to place bounds on the integral $J$,
which can be done as follows: First write 
\be
J = \gamma^2 \undex I \qquad {\rm where} \qquad 
I = 2 \int_1^x x^2 f f_x dx \, .  
\ee
Note that since the function $f$ is of order unity, 
the order of the integral $J$ is given by 
\be
J = {\cal O} \left( \gamma^2 x^2 \right) 
= {\cal O} \left( \tratio^2 \right)\,.
\ee
Next we integrate by parts to obtain 
\be
I = x^2 f^2 - f_0^2 - 2 \int_1^x x f^2 dx \, .
\ee
The second integral can be written 
\be 
2 \int_1^x x f^2 dx = 2 \langle f^2 \rangle \int_1^x xdx = 
\langle f^2 \rangle (x^2 - 1) \,,
\ee
where we have invoked the mean value theorem. The function 
$f$ varies between turning points so that $f_1 \le f \le f_2$. 
We thus have bounds
\be
I \le x^2 (f_2^2 - f_1^2) + f_1^2 - f_0^2 
\le x^2 (f_2^2 - f_1^2) \,,
\ee
and 
\be
I \ge x^2 (f_1^2 - f_2^2) + f_2^2 - f_0^2  
\ge x^2 (f_1^2 - f_2^2) \,. 
\ee
As a result, we have the bound 
\be
|I| \le x^2 (f_2^2 - f_1^2) = x^2 (f_2 + f_1) 
(f_2 - f_1) = x^2 4 a_\ast^2 e_\ast \,. 
\ee
We thus obtain the desired bound on $J$, i.e., 
\be
|J| \le \undex \gamma^2 x^2 (f_2 + f_1) (f_2 - f_1) = 
\gamma^2 x^2 4 \undex a_\ast^2 e_\ast \,. 
\ee
In order to evaluate this bound, we need expressions for the turning
points $f_1$ and $f_2$, or, equivalently, the effective semimajor axis
$a_\ast$ and eccentricity $e_\ast$. As derived in the text, we have
approximations for these quantities, where these expressions are exact
in the limit $J \to 0$.

\label{lastpage} 

\end{document}